# Non-trivial Topological Surface States Regulation of 1T-$OsCoTe_2$ Enables Selective C—C Coupling for Highly Efficient Photochemical $CO_2$ Reduction Toward $C_{2+}$ hydrocarbons

Kangwang Wang[1], Mingjie Wu[2], Peifeng Yu[1], Hector F. Garces[3], Ying Liang[4], Longfu Li[1], Lingyong Zeng[1], Kuan Li[1], Chao Zhang[1], Kai Yan[5,*], Huixia Luo[1,*]

1 School of Materials Science and Engineering, State Key Laboratory of Optoelectronic Materials and Technologies, Guangdong Provincial Key Laboratory of Magnetoelectric Physics and Devices, Key Lab of Polymer Composite & Functional Materials, Sun Yat-sen University, Guangzhou 510275, China

2 State Key Laboratory of New Textile Materials and Advanced Processing Technologies, Wuhan Textile University, Wuhan 430200, China

3 School of Engineering, Brown University, 182 Hope Street, Providence, USA.

4 The Basic Course Department, Guangzhou Maritime University, Guangzhou 510800, China

5 School of Environmental Science and Engineering, Sun Yat-sen University, Guangzhou 510275, China

* Corresponding author. Email: yank9@mail.sysu.edu.cn (K. Yan); luohx7@mail.sysu.edu.cn (H. X. Luo)

## Abstract

Despite ongoing research, the rational design of non-trivial topological semimetal surface states for the selective photocatalytic $CO_2$ conversion into valuable products remains full of challenges. Herein, we present the synthesis of lattice-matched 1T-$OsCoTe_2$ for the photoreduction upgrading of $CO_2$ to tri-carbon ($C_3$) alkane (propane, $C_3H_8$) by the integration of experimental work and theory prediction/calculation. Experimental studies suggested a high electron-based selectivity of 71.2% for $C_3H_8$ and an internal quantum efficiency of 54.6% at 380 nm. In-situ X-ray photoelectron spectroscopy and X-ray absorption fine structure spectroscopy demonstrated that Co and Os atoms coordinated with Te atoms enable an efficient Os—Te—Co electron transfer to activate the generation of $*CH_3$, *CHOCO and $*CH_2OCOCO$. Density functional theory calculations further confirmed Os—Te—Co electron bridging on the improved $CO_2$ conversion kinetics. To our knowledge, this is the first report suggesting the role of Os atoms in accelerating the photocatalytic $CO_2$ conversion activity of the topological semimetal 1T-$OsCoTe_2$.

**Keywords** Topological surface states, electronic properties, $CO_2$ conversion, $C_{2+}$ hydrocarbons, C—C coupling

## 1. Introduction

The surface states of catalysts, in which charge transport and reactant conversion occur simultaneously, plays a central role compared to the bulk state in determining its catalytic performance [1,2]. However, the surface states of catalysts are easily affected and disrupted by the adsorption and desorption of intermediates, fluctuations of surface defects, and potential changes during photoelectrochemical reactions. As time goes on, this dilemma can be skillfully resolved via establishing non-trivial topological surface states (TSSs), which are energy band structures protected by the bulk symmetry. Under this unique protection, the TSSs are unaffected by surface modifications and multiphase catalytic defects, enhancing the activity and stability of the catalysts [3,4]. Moreover, the energy dispersion relation of the non-trivial TSSs results in extremely high carrier mobility and an exotic partial density of states (PDOS), which is crucial for promoting fast catalytic reactions [5,6].

1T-$CoTe_2$, a typical type-II Dirac semimetal of transition metal-sulfide compounds, has been at the frontier of electrocatalytic research due to its diverse phase structure and ultrahigh carrier mobility [7,8]. More specifically, $Co^{2+}$ sites play a crucial role in improving the selectivity of multi-carbon ($C_{2+}$) hydrocarbons, which can not only accelerate the adsorption of *CO and promote the C—C coupling reaction but also stabilize the critical intermediates for generation of $C_{2+}$ hydrocarbons [9,10]. Unfortunately, it remains a challenge for achieving satisfied products selectivity and internal quantum efficiency (*IQE*) due to the complex multiple-electron reduction in the $CO_2$ reduction reaction ($CO_2$RR) process. The introduction of another doping metal is an effective way to regulate and improve the stability of $Co^{2+}$, while the difference in atomic radii between the heteroatom and the Co atom causes lattice distortion, which affects the *d*-band center of the catalyst and thus changes the binding ability of the reaction intermediates on the catalyst surface [11]. More importantly, the doping metal also affects the activity of the competitive hydrogen evolution reaction (HER) [12]. Therefore, modulation of *CO binding energy and suppression of HER are essential objectives in developing metal-atom doped Co-based catalysts.

Based on the considerations mentioned above, integrating the advantages of a topological semimetal (1T-$CoTe_2$) unit with robust surface states and a normal semiconductor (2H-$OsTe_2$) unit into one alliance would be of great interest to optimize their potential for high-performance $CO_2$RR. In this paper, for the first time, we present a detailed investigation of the electronic structure and catalytic mechanism of 1T-$OsCoTe_2$ via combining first-principles calculations with in-situ X-ray photoelectron spectroscopy (XPS) and X-ray absorption fine structure spectroscopy (XAFS) experiments. The investigations demonstrate that the synergy between Co and Os metal sites is crucial for enhanced photocatalytic $CO_2$RR performance. Specifically, Co atoms act as the active catalytic sites, while adjacent Os atoms serve as promoters, and Te atoms function as the transport bridge facilitating electron flow. Density functional theory (DFT) calculations corroborate the strong orbital coupling between Os and Co atoms, causing the *d*-band structure with improved photoactivity. In-situ diffuse reflectance-infrared Fourier-transform spectroscopy (DRIFTS) combined with theory calculations reveal that the synergistic effect of Os and Co metal sites as well their Te electron bridging not only reduces the reaction barrier for the formation of *CHOCO and *$CH_2$OCOCO, but also retards undesired HER, synergistically boosting $CO_2$ conversion to propane (tri-carbon ($C_3$) alkane, $C_3H_8$).

## 2. Experiment section

### 2.1 Materials and Reagents

Tellurium powder (Te, more than 200 mesh) was purchased from Shanghai Aladdin Biochemical Technology Co., Ltd. Cobalt powder (Co, 99.99%) and Osmium powder (Os, 99.99%) were used without further purification. 5,5'-dimethyl-1-pyrroline N-oxide (DMPO) furnished by Shantou Xilong Chemical Co., Ltd. in China was used to trap the generated *$CH_3$ or *OH radicals in the reaction. All of the reagents used in our experiments have analytical grade purity and were used without further purification. De-ionized water ($H_2O$) was obtained from a Millipore system.

### 2.2 Preparation of 1T-$OsCoTe_2$ and 2H-$OsCoTe_2$

To synthesize polycrystalline samples of 1T-$OsCoTe_2$ and 2H-$OsCoTe_2$, stoichiometric amounts of the constituent elements were weighted and grounded with mortar and pestle under inert atmosphere (glovebox). The reaction mixtures were then loaded into quartz tubes and subsequently sealed under vacuum ($10^{-3}$ mbar). The synthesis of 1T-$OsCoTe_2$ was performed via heating the evacuated quartz tubes at 1000 ℃ for 4 h with a heating rate of 5 ℃·$min^{-1}$, subsequently the temperature was decrease to 800 ℃ at 5 ℃·$min^{-1}$ and kept for 24 h. Finally, the mixtures were cooled to room temperature at 0.5 ℃·$min^{-1}$. 2H-$OsCoTe_2$ was synthesized by slowly heating the ampoule with the reaction mixture to 1000 ℃, keeping it at 1000 ℃ for 5 days, and cooling the ampule slowly to room temperature over 1 day. In order to ensure the homogeneity of the samples, the obtained mixtures were reground and heated again at 1000 ℃ for 4 days. After cooling to room temperature, 2H-$OsCoTe_2$ was obtained.

The 1T-$CoTe_2$ and 2H-$CoTe_2$ were prepared following the above synthesis steps of 1T-$OsCoTe_2$ and 2H-$OsCoTe_2$ without adding Os powder. The 2H-$OsTe_2$ was prepared following the above synthesis steps of 2H-$OsCoTe_2$ without adding Co powder. It is worth to mention here that, to gain deeper insight of Te electron bridging characteristics, the pure OsCo alloys were prepared following the above synthesis steps of 1T-$OsCoTe_2$ without adding Te powder.

### 2.3 Photocatalytic $CO_2$RR measurements

Photocatalytic $CO_2$RR measurement was conducted in an online trace gas analysis system with a gas chromatography-mass spectrometry (GC-MS, Agilent GC/MS-7000D) and $^1$H nuclear magnetic resonance ($^1$H NMR) spectroscopy. Firstly, 50 mg of photocatalyst and 100 mL of aqueous acetonitrile (MeCN) solution [13] were added in the Pyrex glass reaction cell with sonication. Then, the reactor system was then filled with pure $CO_2$ gas (80 kPa) following complete evacuation. After adsorption equilibrium, the reactor system was put under a simulated light source which was provided by a 300 W Xe lamp (Beijing Perfectlight,

Microsolar 300, 100 mW $cm^{-2}$) with a UVCUT-420 nm filter to cut-off the light with wavelengths lower than 420 nm (i.e. $\lambda > 420$ nm), and the chilled water controlled the system temperature. Finally, the gas and liquid produced were analyzed every 1 h through GC-MS and $^1H$ NMR spectroscopy (Bruker AVIII HD 600), respectively. Notably, for the $^1H$ NMR measurement, 0.5 mL of the product solution was mixed with 0.1 mL of $D_2O$ and 10 μL of dimethylsulfoxide (DMSO, as the internal standard). In additon, all the experiments were repeated at least 3 times in parallel to obtain an average value and error bars indicate standard deviations.

### 2.4 Characterization

Powder X-ray diffraction (PXRD) patterns were collected on an X-ray diffractometer (X' Pert3 Powder) equipped with a Kratky block-collimation system at 25.0 ± 0.1 ℃ and a sealed-tube X-ray generator with the Cu target operating at 45 kV and 40 mA. Inductively coupled plasma optical emission spectrometer (ICP-OES) was tested on an Agilent 7500cx instrument with an attached laser ablation system. UV-Vis-NIR diffused reflectance spectra (DRS) was collected on a spectrophotometer (UV-3100, Shimadzu, Japan) with $BaSO_4$ as the background holder. Steady-state photoluminescence (PL) and time-resolved PL (TRPL) measurements were performed on a Hamamatsu instrument (Hamamatsu C5680, Japan), with an excitation wavelength of 320 nm. $CO_2$ adsorption isotherms were obtained at 25 ℃ via using a Quantachrome Autosorb-iQ adsorption analyzer after the degassing process at 200 ℃ for 12 h.

### 2.5 Computational method of charge distribution

All of the calculations are performed in the framework of the spin-polarized density functional theory with the projector augmented plane-wave method, as implemented in the Materials Studio 2019 and Vienna ab initio simulation package (VASP) [14]. The generalized gradient approximation (GGA) proposed by Perdew, Burke, and Ernzerhof is selected for the exchange-correlation potential [15,16]. The van der Waals interaction was taken into account using DFT-D3 method with Becke-Jonson damping dispersion correction [17,18]. The cut-off energy for plane wave was set to 400 eV. The energy criterion was set to $10^{-5}$ eV in iterative solution of the Kohn-Sham equation. The Brillouin-zone sampling was conducted using Monkhorst-Pack (MP) grids of special points with the separation of 0.04 $Å^{-1}$. All the structures are relaxed until the residual forces on the atoms have declined to less than 0.02 eV $Å^{-1}$. A Gaussian smearing of 0.05 eV was applied to speed up self-consistent field iteration. Data analysis and visualization are carried out with the help of VASPKIT, Materials Studio 2019, and VESTA. DFT + $U$ approach [19] with $U$ = 3.3 eV was considered to evaluate the influence of strongly correlated $d$ electrons of Co and Os on the calculated free energies. Adsorption energy ([ads]) was defined as $E_{[ads]} = E_{[adsorbate/substrate]} - E_{[substrate]} - E_{[adsorbate]}$, where $E_{[substrate]}$ is the total energy of substrate, $E_{[adsorbate/substrate]}$ is the total energy when carbon monoxide (CO), methane ($CH_4$), ethylene ($C_2H_4$) or $C_3H_8$ is adsorbed on the substrate surface, $E_{[adsorbate]}$ refers to the energy of CO, $CH_4$, $C_2H_4$ or $C_3H_8$. The Gibbs free energy change ($\Delta G$) was defined as $\Delta G = \Delta E + \Delta E_{ZPE} - T\Delta S$, where $\Delta E$ is the energy difference between the reactants and product obtained through DFT calculations. $\Delta E_{ZPE}$ and $\Delta S$ are the changes in the zero-point energies (ZPE) and entropy. $T$ represents the temperature and was set as 298.15 K.

## 3. Result and discussion

### 3.1 Structural Characterization

Based on transmission electron microscopy (TEM) image, 1T-$OsCoTe_2$ demonstrates a typical bulk-like structure and possesses abundant pores (Fig. 1a). Notably, energy-dispersive X-ray spectroscope (EDS) mapping indicates that Co elements on 1T-$OsCoTe_2$ are distributed around Os elements, and there are high-density bright spots on Te elements, implying atomically dispersed Co and Os species (Fig. 1b). Meanwhile, high-resolution TEM (HRTEM) image shows that the lattice spacings of 3.64 and 1.93 Å (Fig. 1 e) correspond to (111) plane of 1T-$CoTe_2$ and (311) plane of 2H-$OsTe_2$, respectively (Fig. 1 c). From the results, the phases of 1T-$CoTe_2$ and 2H-$OsTe_2$ belong to P-3m1 and Pa-3 space groups, respectively. This intuitive result provides direct evidence for successfully constructing 2H-$OsTe_2$ and 1T-$CoTe_2$ composite structures. To further confirm, an aberration-corrected high-angle annular dark-field scanning TEM (AC HAADF-STEM) image is employed and displayed two kinds of points with different brightness and colors through pseudo-color transformation (Fig. 1f and g). The HAADF-STEM image of 1T-$OsCoTe_2$ reveals a crystal structure resembling that of 1T-$CoTe_2$ (P-3m1) and 2H-$OsTe_2$ (Pa-3) (Fig. S1), with no discernible Os crystalline phase observed, potentially attributable to the low and uniform distribution of Os species (Fig. 1g). By comparing the magnified HAADF-STEM images, they match well with the lattice models after XRD optimization (Fig. S2). Selected regional electron diffraction (SAED) pattern shows diffraction rings and spots, suggesting polycrystalline and monocrystalline composite properties of 1T-$OsCoTe_2$ (Fig. 1h). As an overall picture, the chemical composition of 1T-$OsCoTe_2$ is quantitatively determined by ICP-OES results (Fig. S3a), which is similar to the chemical formula of 1T-$OsCoTe_2$. Meanwhile, diffraction rings of 2H-$OsCoTe_2$ can be observed in the SAED pattern, representing the polycrystalline property of 2H-$OsCoTe_2$ (Fig. 1d and S3b–e).

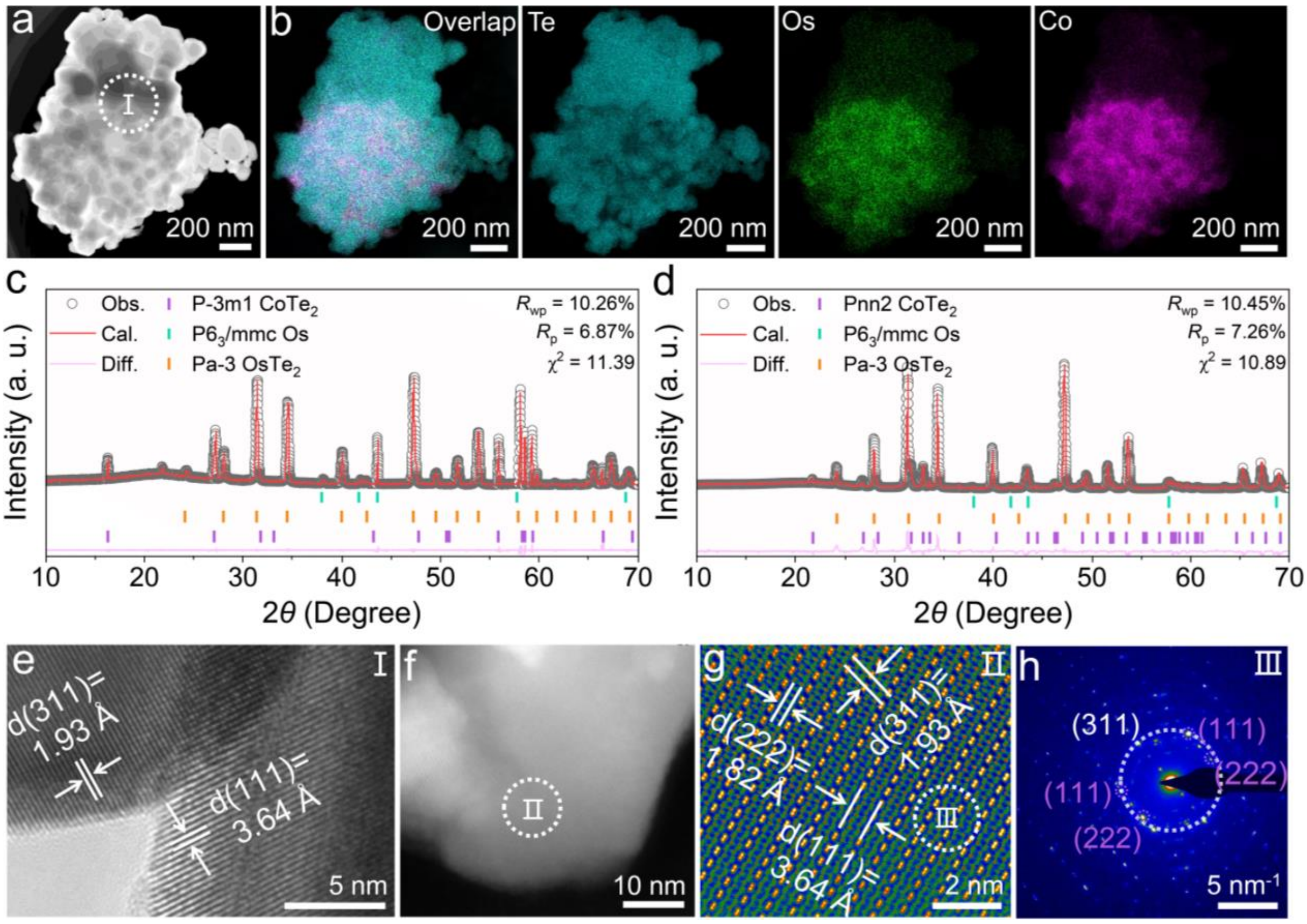

**Fig. 1.** (a) TEM and (b) EDS elemental mapping of 1T-$OsCoTe_2$, respectively. XRD profiles and Rietveld refinement results of (c) 1T-$OsCoTe_2$ and (d) 2H-$OsCoTe_2$, respectively. (e) HRTEM image of 1T-$OsCoTe_2$. (f,g) HAADF-STEM images and (h) SAED pattern of 1T-$OsCoTe_2$, respectively.

### 3.2 Electronic Structure Analysis

The electronic structure information of 1T-$OsCoTe_2$ and 2H-$OsCoTe_2$ were compared via in-situ XPS spectra (Fig. 2 and S4). After 20 min of illumination, the binding energy of Co 2*p* shifts to the right via 1.97 eV, and then returns to the "dark-state" after 10 min of light-off, suggesting electronic gain of 1T-$CoTe_2$ upon light illumination (Fig. 2a). This is mainly due to the electrons in the conduction band of 2H-$OsTe_2$ being transferred to Co 2*p* orbitals on 1T-$CoTe_2$ through a built-in electric field force [20,21], and the binding energy change of Os 4*f* on 2H-$OsTe_2$ further confirms the direction of electron migration (Fig. 2b). Theoretically, the binding energy shift of Te 3*d* should be opposite to that of Os 4*f* on 2H-$OsTe_2$. Nevertheless, this is not the case, and the binding energy of Te 3*d* shifts to the right by 3.02 eV after 20 min of light-on, and then to the left via 1.47 eV after 40 min of light-on. In sharp contrast, the binding energy of Te 3*d* does not completely return to the "dark-state" after 10 min of light-off (Fig. 2c) [22], which indicates that there are two electron transport processes in 1T-$OsCoTe_2$ under light radiation. The first process is a direct electron transfer from Os atoms to Te atoms (i.e., causing to a decrease in the binding energy of Te 3*d*). The second process is an indirect electron transfer from Te atoms to Co atoms driven by an interfacial electric field (i.e., leading to an increase in the binding energy of Te 3*d*). The binding energy core-level peak for Te 3*d* does not shift, confirming that the two processes are balanced. Interestingly, the small changes in Os 4*f* binding energy before and after illumination disclose that the first electron transport process is the dominant one. The bulk electronic band structures and PDOS of 1T-$OsCoTe_2$, 2H-$OsCoTe_2$, 1T-$CoTe_2$, and 2H-$CoTe_2$ without spin-orbit coupling (SOC) were calculated as shown in Fig S5 and S6. The few crossings and low PDOS at the Fermi level ($E_F$) of 1T-$CoTe_2$ display its semimetallic properties. However, the PDOS at the Fermi surface in 1T-$OsCoTe_2$ is clearly higher than that in 2H-$OsCoTe_2$, indicating faster charge transfer and lower resistance of 1T-$OsCoTe_2$ than 2H-$OsCoTe_2$, which is in accordance with the transient photocurrents. In addition, the bulk electronic band structure (Fig S5a and b) denotes that the conduction and valence bands located near the $E_F$ intersecting along G−L direction forms Dirac node, suggesting that 1T-$CoTe_2$ possess ultra-high conductivity and mobility, which is a general feature of Dirac semimetal [6,7,23]. Thus, topologically protected surface states are accommodated at the 1T-$CoTe_2$ (001) facet, which appear in the bandgap at the Γ point, as demonstrated in Fig S6a and c. Notably, the presence of TSSs was found in the surface energy bands of 1T-$CoTe_2$ (001) with surface state energy levels near the $E_F$ (~0.01 eV higher) (Fig S5b and e–i) [24], which means that non-trivial TSSs can provide additional electrons and facilitate the electrons transfer of Os—Te—Co electron bridging from the surface of 1T-$OsCoTe_2$ to the adsorbed $CO_2$ molecules, which is conducive to $CO_2$RR.

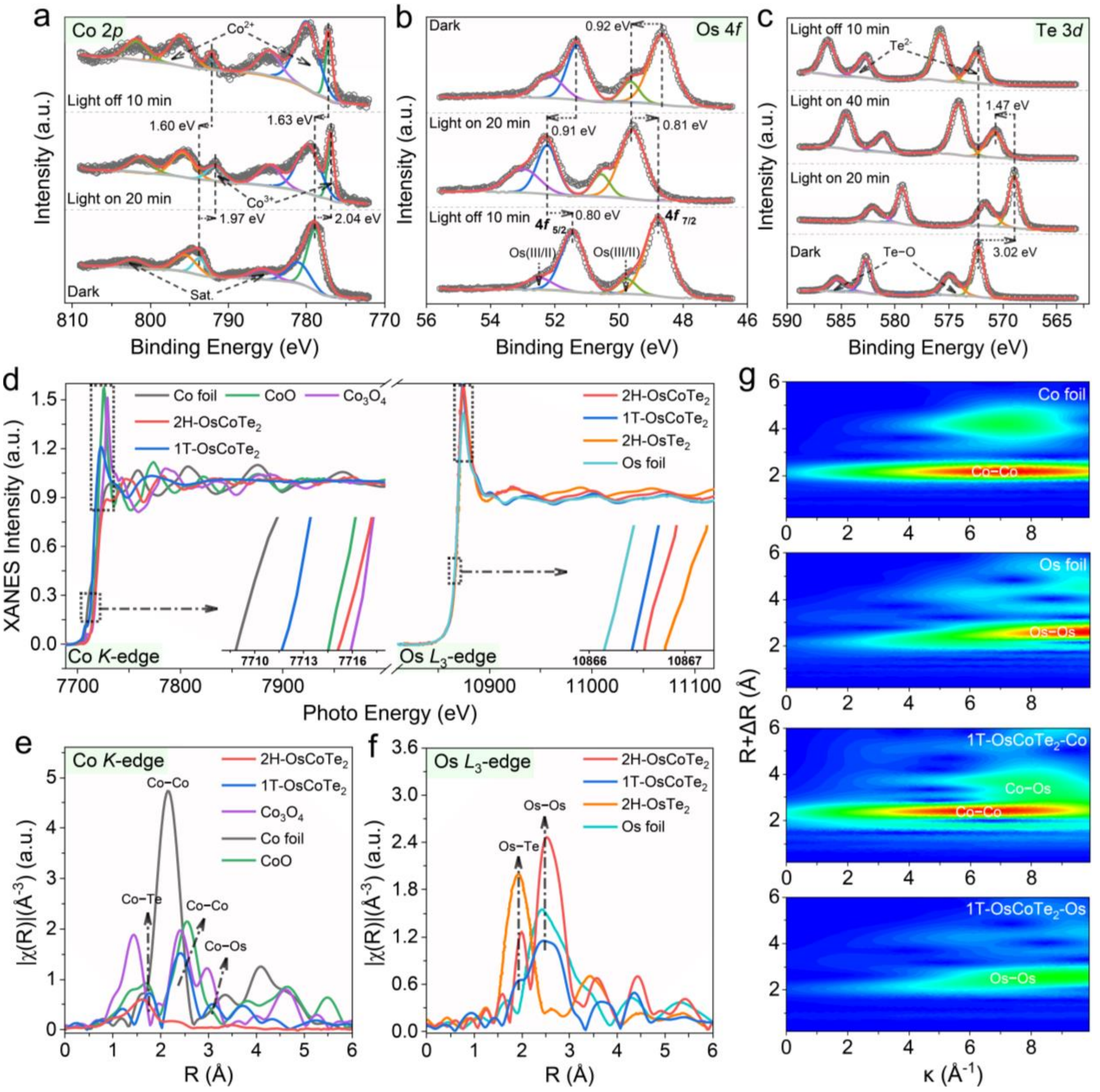

**Fig. 2.** In-situ XPS (a) Co 2*p*, (b) Os 4*f*, and (c) Te 3*d* spectra of 1T-$OsCoTe_2$. (d) Co *K*-edge and Os $L_3$-edge XANES spectra of different samples. (e) Co *K*-edge and (f) Os $L_3$-edge EXAFS in R-space. (g) WT spectra for the $\kappa^3$-weighted $\chi(\kappa)$ *K*-edge EXAFS signals of Co foil, Os foil, Co *K*-edge, and Os $L_3$-edge of 1T-$OsCoTe_2$.

To systematically explore the active sites of the photocatalysts as well as the coordination environments of Os and Co atoms, XAFS spectroscopy was implemented for 1T-$OsCoTe_2$ and 2H-$OsCoTe_2$. X-ray absorption near-edge structures (XANES) of Co *K*-edge indicate that the absorption edge of 2H-$OsCoTe_2$ is located between CoO and $Co_3O_4$, meaning that the valence of Co element is between +2 and +3 [25]. The enlarged profile clarifies that the valence state of the Co element in 2H-$OsCoTe_2$ is close to + 2. Meanwhile, XANES spectra of Os $L_3$-edge show that the positions of the absorption threshold of 1T-$OsCoTe_2$ and 2H-$OsCoTe_2$ are situated in Os foil and 2H-$OsTe_2$ (Fig. 2c), which reveals that the valences of Os element are between 0 and +2 [26]. Extraordinarily, the intensity of the white-line peak of Co *K*-edge XANES spectrum of 1T-$OsCoTe_2$ is lower than that of CoO, indicating a decrease in the PDOS and an increase in the electron filling of the *d*-band for the unoccupied state of Co 3*d* near the Fermi energy level. This result suggests that electrons are injected into the Co 3*d* orbitals after introducing 2H-$OsTe_2$ [27,28]. Contrastingly, the white-line peak intensity of Os $L_3$-edge XANES spectrum of 1T-$OsCoTe_2$ is higher than that of 2H-$OsTe_2$. Moreover, the intensity of white-line peak and absorption edge of 1T-$OsCoTe_2$ are significantly different from that of Os foil, while it is extremely close to that of 2H-$OsTe_2$ (Fig. 2c), which suggests that the average Os valence state of 1T-$OsCoTe_2$ is similar to that of 2H-$OsTe_2$, agreement with XPS results [26,29]. It is estimated that Co atoms sequentially form electron accumulation layers by swiping electrons from the surrounding Os atoms through Co—Te bonds. Significantly, the charge density around the Co atoms increases, further demonstrating the electron-donating nature of Os—Te bonds. As displayed by the PDOS in Fig. S6d and e, the higher electron density around the Co atoms modulates the 3*d* orbital structure, leading to a positive shift in the center of the *d*-band. The resulting electron-deficient Os atoms bring the whole 1T-$OsCoTe_2$ close to a relatively high valence state, which can be further verified from the results of Os $L_3$-edge $k^2$-weighted oscillation (Fig. S6f). The first significant peak in EXAFS spectra of 1T-$OsCoTe_2$ (Fig. 2d) denotes that Co atoms are coordinated with Te atoms nearest neighboring shell locates at ~1.76 Å, referring to the peak of Co—Te bonds in 1T-$CoTe_2$ [30]. Simultaneously, the first significant peak can be attributed to Os—Te bonds that Os atoms are coordinated through Te atoms nearest neighboring shell at ~1.93 Å, which is well consistent with that of 2H-$OsTe_2$, indicating that Os atoms interact with Te atoms through Os—Te bonds (Fig. 2e). The detailed coordination environments of Co and Os atoms were further investigated by Fourier transform of extended X-ray absorption fine structures (EXAFS) and wavelet transforms (WT) spectra (Fig. 2f and g). The second prominent peak at ~2.59 Å in Os $L_3$-edge EXAFS spectrum of 1T-$OsCoTe_2$ can be attributed to Co—Os coordination at the first shell, which is similar to 2H-$OsCoTe_2$ and different from that of Os—Te coordination. Of note, no significant peaks such as Os—Os (~2.68 Å) are detected on the longer backscattering paths ((Fig. S7 and Table S1), indicating Os species in 1T-$OsCoTe_2$ and 2H-$OsCoTe_2$ exist in a single atomic state [26,31], which is consistent with XRD fitting results. Collectively, after the formation of 1T-$OsCoTe_2$, the number of electrons on Co atoms increases and the number of electrons on Os atoms decreases, and there may be transferred from Os to Co via Os—Te—Co electron bridging.

### 3.3 Photocatalytic $CO_2$RR Performance

The $CO_2$ photoreduction products were measured in MeCN aqueous solution (MeCN:$H_2O$ = 6:1, Fig. S8a) to evaluate $CO_2$RR performance of the catalyst [13,32,33]. 50 mg of photocatalyst loading was adopted to improve $C_3H_8$ selectivity (Fig. S8b). A similar photocatalyst dosage was often used in previous photocatalytic works. The gaseous and liquid reduction products were analyzed and quantified via GC-MS and $^1$H NMR spectroscopy, respectively (Fig. 3a and S9a–d). CO and $CH_4$ are mainly detected on 1T-$CoTe_2$, 2H-$OsTe_2$, and OsCo with yields of 33.74, 23.61, and 11.3 μmol $g^{-1}$ $h^{-1}$, respectively (Fig. 3a). 1T-$OsCoTe_2$ exhibits significant yields of $C_2H_4$ (3.26 μmol $g^{-1}$ $h^{-1}$) and $C_3H_8$ (10.28 μmol $g^{-1}$ $h^{-1}$), together with a small amount of $CH_4$ (1.78 μmol $g^{-1}$ $h^{-1}$), in addition to CO (14.91 μmol $g^{-1}$ $h^{-1}$) (Fig. 3b and c), exhibiting an exceptionally strong C—C coupling capability [34]. The $C_2H_4$ and $C_3H_8$ yields of 2H-$OsCoTe_2$ are reduced to 2.27 and 7.72 μmol $g^{-1}$ $h^{-1}$, indicating that the non-trivial TSSs of 1T-$OsCoTe_2$ is conducive to the improvement of $CO_2$ activity through enhancing the adsorption of intermediates, promoting carrier mobility and reducing the activation energy of reactants [4,5]. Strikingly, an $IQE_{cr}$ of 1T-$OsCoTe_2$ at the wavelength of 380, 400, 420, 440, 460, 480, 500, and 520 nm is 54.6%, 25.63%, 10.84%, 9.49%, 7.12%,

5.25%, 2.80%, and 1.27%, respectively (Table S2) [35, 36]. A high product-based selectivity of $C_3H_8$ of 34.0% is achieved (71.20% of the electron-based selectivity), and that of total $C_{2+}$ hydrocarbons is as high as 44.8% (84.7% of the electron-based selectivity) (Fig. 3b) [37]. The catalytic activity of 1T-$OsCoTe_2$ and 2H-$OsCoTe_2$ is competitive with those of well-known photocatalysts under strict conditions (Table S3). Herein, we further tested the catalytic performance of 1T-$OsCoTe_2$ and 2H-$OsCoTe_2$ in pure $H_2O$ (with $H_2O$ as the electron donor), and the results shows that the generation rate of all carbon-based products decreased (Fig. S10a). Moreover, the $C_{2+}$ hydrocarbon selectivity of 1T-$OsCoTe_2$ becomes lower in pure $H_2O$ (Fig. S10b). Interestingly, no $C_{2+}$ hydrocarbons of 2H-$OsCoTe_2$ are detected in pure $H_2O$, but more $H_2$ is produced with a formation rate of 3.84 μmol $g^{-1}$ $h^{-1}$ (10.27% of the electron-based selectivity) (Fig. S8c and d). The improvement of the catalytic activity and selectivity may be attributed to the high solubility of $CO_2$ in MeCN, which significantly increases the local accessibility of $CO_2$, thus enhancing the contact between 1T-$OsCoTe_2$ and $CO_2$, and promoting C—C coupling [34]. Enlightened via the analyses described above, the effective inhibition of CO generation and increased $C_{2+}$ hydrocarbons production on 1T-$OsCoTe_2$ and 2H-$OsCoTe_2$ confirm that the formation of $C_{2+}$ hydrocarbons are probably owing to the C—C coupling depletion of *CO.

The gas production rate of 1T-$OsCoTe_2$ under light irradiation was measured by cyclic tests, as shown in Fig. 3d and S11a–c, which is pivotal for evaluating the stability of the photocatalyst. The distinctive $CO_2$ photoreduction activity and high selectivity of $C_3H_8$ and total $C_{2+}$ hydrocarbons are still well maintained after eight cycling tests of 40 h. This is further supported through XRD (S11d), XPS (Fig. S12) and XAFS (Fig. S13) spectra of 1T-$OsCoTe_2$ after the stability tests. All the post-reaction characterizations indicate that 1T-$OsCoTe_2$ has the excellent capability and stability to efficiently achieve the photoreduction of $CO_2$ to $C_{2+}$ hydrocarbons. Moreover, the isotropic experiment was further carried out by employing $^{13}CO_2$ as a reactant to ascertain the carbon source of the products. $^{13}CO_2$ isotope labeling experiment and the time profile of the relative abundance of $^{13}C$ labeled products confirm that the carbon source for CO(g) and other hydrocarbon products originates from the input $CO_2$(g) (Fig. S9c and d) [38,39]. To explore the role of Te atoms in 1T-$OsCoTe_2$, OsCo was synthesized under the same conditions. In sharp contrast, OsCo is nearly photo-catalytically incapable toward $CO_2$RR (Fig. 3b), confirming that Os—Te—Co electron bridging of 1T-$OsCoTe_2$ promote photocatalytic activity. To elucidate the change in the valence states and the local structures surrounding Os and Co atoms, XAFS experiments further were performed in detail (Fig. S14). Compared with 1T-$OsCOTe_2$, both the valence states of Os and Co atoms shift to lower energy in OsCo. Co *K*-edge XANES curves show that the absorption edge of OsCo is between that of Co foil and 1T-$CoTe_2$ (Fig. S14a), indicating that the valence state of Co atoms in OsCo is between 0 and +2, which may be caused by surface oxidation. Moreover, OsCo exhibits a lower intensity than 1T-$CoTe_2$ (inset in Fig. S14a), implying that the introduction of Te atoms reduces the oxidation state of Co atoms in 1T-$CoTe_2$ [40]. Meanwhile, EXAFS fitting shows that Co atoms in OsCo without the Te electron bridging is coordinated with Os atoms (Fig. S14b–f). As such, 1T-$OsCOTe_2$ with Te electron bridging significantly accelerates photocatalytic $CO_2$ reduction to $C_{2+}$ hydrocarbons.

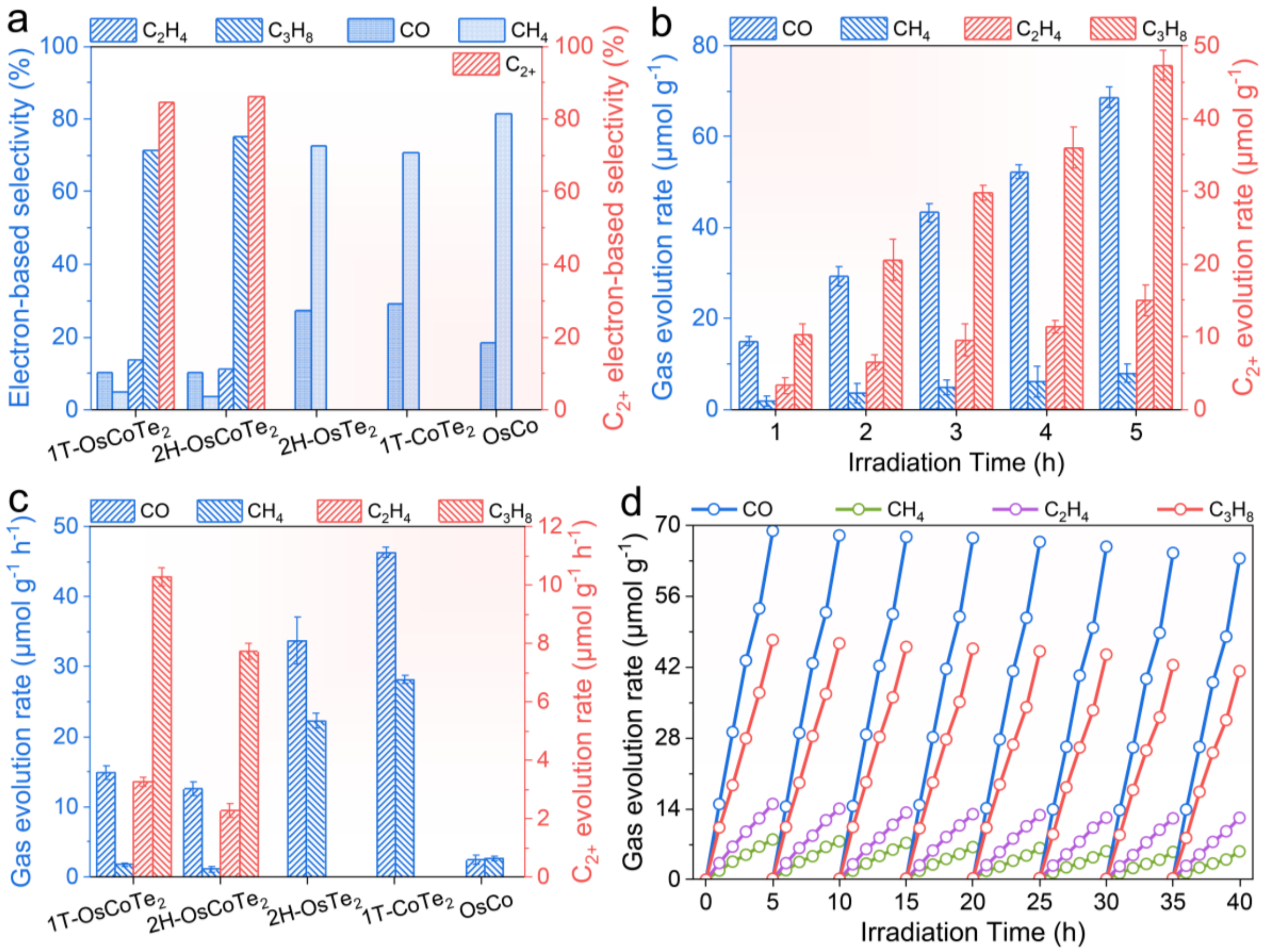

**Fig. 3.** (a) Electron-based selectivity on as-prepared samples in MeCN aqueous solution. (b) Photocatalytic product evolution as a function of light irradiation times on 1T-OsCoTe$_2$ in MeCN aqueous solution. (c) Product formation rates on samples in MeCN aqueous solution. (d) Gas evolution amounts as a function of light irradiation time for 1T-OsCoTe$_2$ over eight cycling tests in MeCN aqueous solution.

### 3.4 Photoelectric Performance Analysis

The absorption edge of 1T-OsCoTe$_2$ displays a blue shift compared with 2H-OsCoTe$_2$ due to the non-trivial TSSs of 1T-OsCoTe$_2$, while the light absorption is enhanced after implantation of 1T-CoTe$_2$ (Fig. S15a). Transient photocurrent measurements confirm the enhanced charge separation and migration efficiency of 1T-OsCoTe$_2$, ascribed to the non-trivial TSSs, which effectively shortens the charge transfer distance from body to surface and lowers charge recombination possibility (Fig. S15b). Theoretically, the fast charge carrier dynamics of 1T-OsCoTe$_2$ is kinetically favorable for the multi-electron reactions of generating $C_{2+}$ hydrocarbons. Therefore, the photogenerated carrier separation efficiency was determined via steady-state PL, TRPL, and transient photocurrents. Dramatic PL quenching occurs on 1T-$OsCoTe_2$ relative to OsCo, implying that the intrinsic radiative recombination of photogenerated carrier in 2H-$OsTe_2$ has been substantially inhibited due to the promoted charge transfer through Os—Te—Co electron bridging (Fig. S15c). Moreover, the TRPL decay spectra recorded at the corresponding steady-state emission peaks illustrate that 1T-$OsCoTe_2$ has a longer average lifetime ($\tau_{avg}$) of charge carriers with respect to 1T-$CoTe_2$ and 2H-$OsTe_2$ (Fig. S15d and Table S5), meaning that there is efficient charge transfer in 1T-$OsCoTe_2$ [41]. As shown in Fig. S15b, 1T-$OsCoTe_2$ is favorable for charge separation from the enhancement of photocurrents. Peculiarly, the transient photocurrents density and PL density of 2H-$OsTe_2$ are close to those of OsCo, illustrating that these factors do not have much influence on the performance difference between 2H-$OsTe_2$ and OsCo [42−44]. In contrast, the non-trivial TSSs of 1T-$OsCoTe_2$ are better than those of 2H-$OsCoTe_2$, and thus should be the reason for the relatively higher performance of 1T-$OsCoTe_2$. Ultrafast femtosecond time-resolved transient absorption (fs-TA) spectroscopy provided evidence for the mechanism of photogenerated carrier transfer. As shown in Fig. 4a and b, TA spectra of 1T-$OsCoTe_2$ and 2H-$OsCoTe_2$ both demonstrate pronounced positive and negative absorption peaks at 440 and 493 nm, belonging to the excited state absorption (ESA) and ground state bleach (GSB), respectively [45,46], in which the GSB peak reflects the relaxation of the excited state. A redshift of GSB peak with pump-probe delay time can be observed in Fig. 4c and d, mainly due to the broad size distribution of Os atoms in 1T-$OsCoTe_2$ and 2H-$OsCoTe_2$, where the smaller Os atoms have the larger band gaps and faster exciton annihilation [47,48]. To deeply analyze the dynamic behavior of charge carriers, GSB and ESA decay kinetics curves of both 1T-$OsCoTe_2$ and 2H-$OsCoTe_2$ are shown in Fig. 4e and f, respectively, which indicate that the kinetic decays of GSB and ESA are significantly suppressed on the probing time scale. The clear negative absorption bands attributed to the GSB peak signal can be observed, which implies charge carrier transfer between Os—Te—Co electron bridging [49], agreeing with the results of in-situ XPS and XAFS.

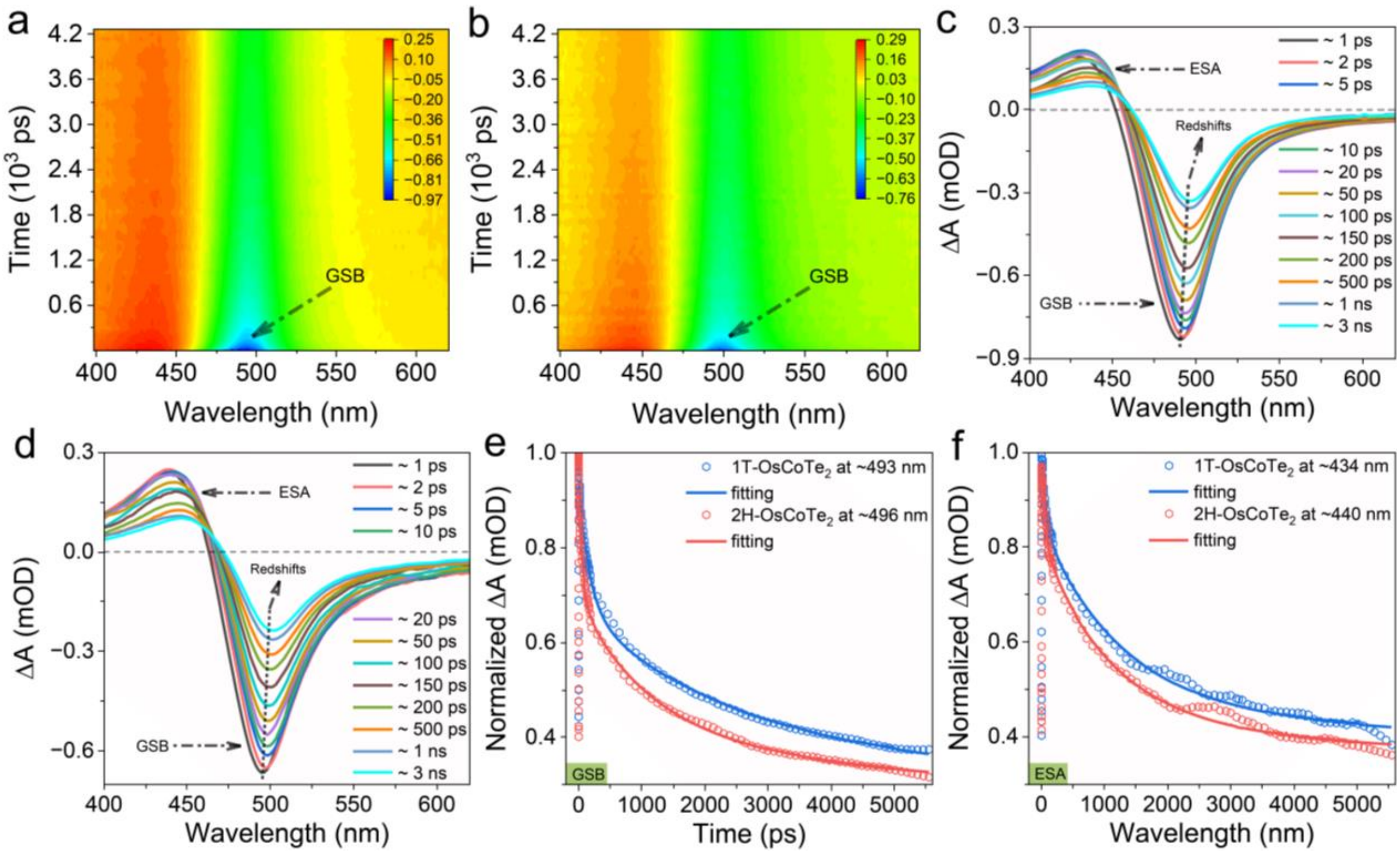

**Fig. 4.** 2D Pseudo-color images of (a) 1T-$OsCoTe_2$ and (b) 2H-$OsCoTe_2$ in ethanol solution after the excitation with a 310 nm laser pulse. The TA spectra of (c) 1T-$OsCoTe_2$ and (d) 2H-$OsCoTe_2$ at different delays time. (e) Comparison of decay kinetics and fitting lines for 1T-$OsCoTe_2$ and 2H-$OsCoTe_2$ taken through the GSB peaks at ~493 and ~496 nm, respectively. (f) Normalized decay kinetics and fitting lines for 1T-$OsCoTe_2$ and 2H-$OsCoTe_2$ taken through the ESA peaks at ~434 and ~440 nm, respectively.

### 3.5 Reaction Mechanism Analysis

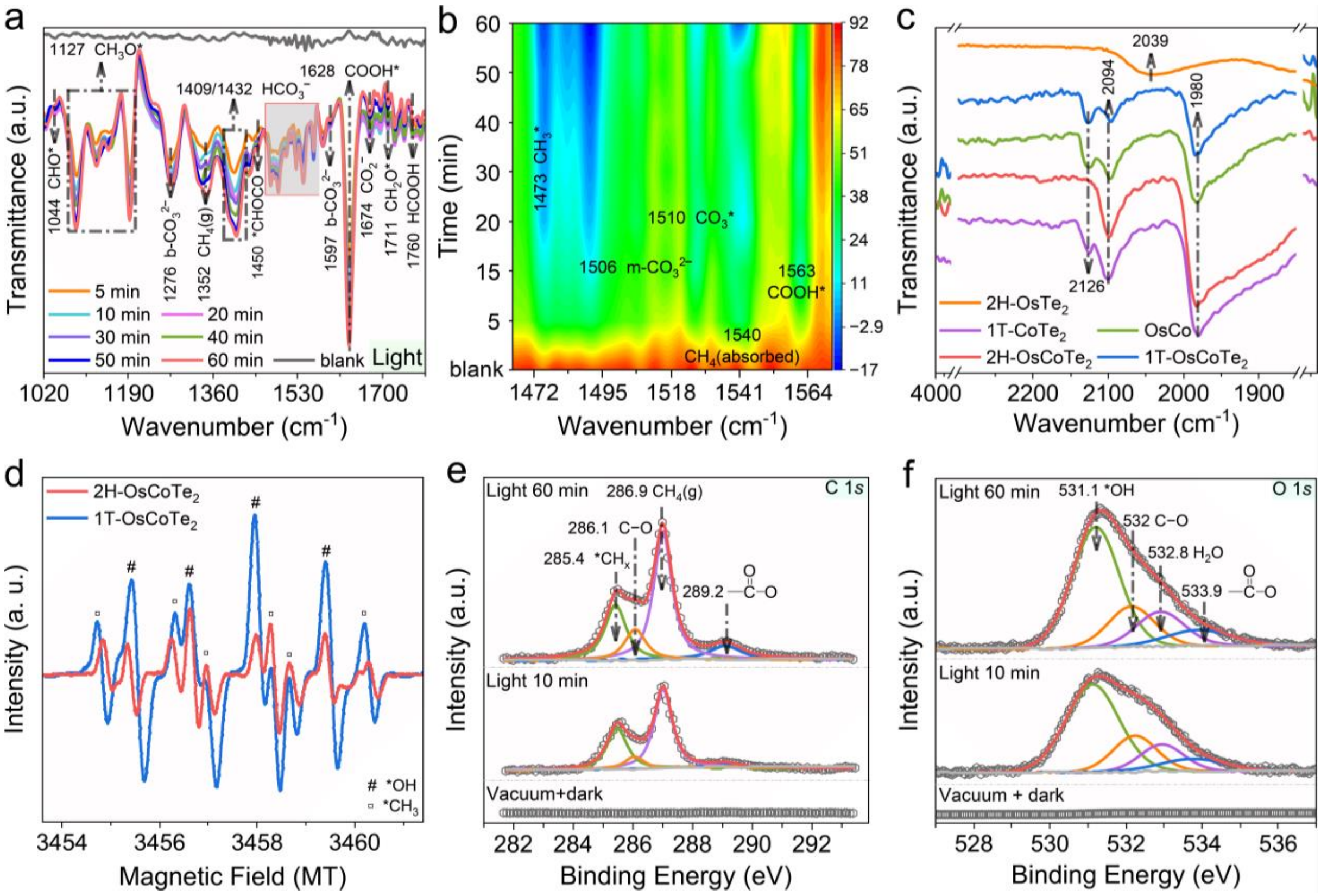

**Fig. 5.** (a,b) In-situ DRIFTS spectra for light-driven $CO_2$ conversion over 1T-$OsCoTe_2$. (c) CO adsorption DRIFTS spectra of samples. (d) In-situ ESR spectra of $CO_2$ + $H_2O$ mixture under light illumination at 298 K in the presence of DMPO over 1T-$OsCoTe_2$ and 2H-$OsCoTe_2$. In-situ NAP-XPS results of high-resolution (e) C 1*s* and (f) O 1*s* spectra over 1T-$OsCoTe_2$.

To understand the mechanism of photocatalytic $CO_2$RR at molecular level, the possible intermediates were performed by in-situ DRIFTS technique. In the spectra (Fig. 5a, b and Fig. S16a), the peaks at 1409/1432 cm$^{-1}$ belong to $HCO_3^-$ [50], while the peaks at 1276/1597 cm$^{-1}$ is assigned to bidentate carbonates (b-$CO_3^{2-}$) [51,52]. In addition, a small amount of *$CH_3$ (1473 cm$^{-1}$), *$CO_3$ (1510 cm$^{-1}$), $CH_3O$* (1127 cm$^{-1}$) [37], CHO* (1044 cm$^{-1}$) and $CH_2O$* (1711 cm$^{-1}$) is observed, resulting from the intermediates to the production of $CH_4$ (1352 cm$^{-1}$) and $C_3H_8$ [52,53]. The characteristic peaks of *COOH at 1563 and 1628 cm$^{-1}$ appear as the most critical intermediates in the photoreduction of $CO_2$-to-CO [13]. The increasing peak at 2130 cm$^{-1}$ may be derived from *CO, a key intermediate in the formation of CO. The peaks at 1980 and 2094 $cm^{-1}$ are attributed to linear- and bridge-type CO adsorption at the Co sites, respectively (Fig. 5c). In addition, a new peak arising at 2126 $cm^{-1}$ over 1T-$OsCoTe_2$ and 1T-$CoTe_2$ is attributed to the linear CO adsorption on the non-trivial TSSs (as mentioned in the introduction), demonstrating the well-formed TSS of 1T-$CoTe_2$ in the samples [5]. Of note, a broad peak at 2039 $cm^{-1}$ is observed, which belongs to bridge-type CO adsorption over 2H-$OsTe_2$ [54], corresponding to the detection of carbonyl (*COOH, *CO, *CHO, and *CHOCO) intermediates via in-situ DRIFTS [55,56]. In-situ electron paramagnetic resonance (ESR) spectroscopy was used to further detect free radicals produced by photoinduced activation of various reactants. As shown in Fig. 5d, *$CH_3$ is detected in addition to *OH when $CH_4$ is injected into an aqueous solution containing 1T-$OsCoTe_2$ and 2H-$OsCoTe_2$ under visible light illumination, *$CH_3$, in addition to *OH, are detected (Fig. S16b and c) [57,58]. In order to further understand $CO_2$RR mechanism through elemental information, carbon and oxygen species on the surface of the material were also monitored by in-situ near ambient pressure XPS (NAP-XPS) measurements. As shown in Fig. 5e and S17, a peak of $CH_4$ (287.0 eV) is clearly detected in C 1*s* XPS spectra (Fig. S17a). The C1*s* peaks of *$CH_x$, C—O, and *COO species occurring at 285.4, 286.1, and 289.2 eV increase with the time evolution upon visible light irradiation. Afterward, it is further verified that $CO_2$ was reduced to produce oxygen-containing compounds via collecting the O 1*s* spectrum in NAP-XPS. The characteristic peaks of *OH (531.1 eV), C—O (531.9 eV), $H_2O$ (532.6 eV), and C=O (533.3 eV) could be resolved under visible light illumination (Fig. 5f). Interestingly, strong *CO signals are observed with in-situ DRIFTS and NAP-XPS, but extremely little CO is detected on 1T-$OsCoTe_2$ and 2H-$OsCoTe_2$. Indeed, Fig. 5d already reflects the relatively strong adsorption of CO on the Co active sites, allowing *CO to couple to other intermediates before desorption [59,60]. Taken together, the presence of *$CH_x$, C—O, and *COO was confirmed by a series of in-situ tests, indicating that the co-adsorption reaction of $CO_2$ and $H_2O$ on 1T-$OsCoTe_2$ can generate a variety of surface carbonaceous intermediates (e.g., methyl and carbonyl) [54] and further $C_{2+}$ hydrocarbons. The experimental results have revealed the critical role of 1T-$OsCoTe_2$ in $CH_4$-to-$CH_4$/$C_2H_4$/$C_3H_8$ conversion. Fig. 6a and b illustrate the proposed reaction pathway. $CO_2$ prefers to be activated at Os sites in the presence of $H^+$ and form Os—$OCH_3$. The *$CH_3$ can be gradually converted to Os—O through combining Te atoms from Os—Te unit and the hydrogenation via $H^+$ (Fig. 6b). Simultaneously, the $C_1$—$C_1$ coupling between *CHO and *CO generates Co—CHOCO at Co—Te unit interface, and the further hydrolysis of Co—$OCH_2CO$ gives $C_2H_4$ as a product. The Os sites are also active for Os—CO generation by directly oxidizing CO on Os—Te unit (Fig. 6a). However, the *$CH_3$ formed on Os—Te unit can hardly approach *CO on Co—Te unit, so the surplus *$CH_3$ would evolve into $CH_4$ (Fig. 6b). Moreover, the formation of a five-membered ring between *$CH_2OCOCO$ and Co active sites significantly mitigates the accumulation of localized electrons and weakens inter- and intramolecular electrostatic repulsion (Fig. 6a). The electronic and non-trivial TSSs, as well geometrical effects of 1T-$OsCoTe_2$ may have jointly stabilized critical *$C_{2+}$ intermediates and lowered their adsorption energy levels, thus facilitating C—C coupling [61−63].

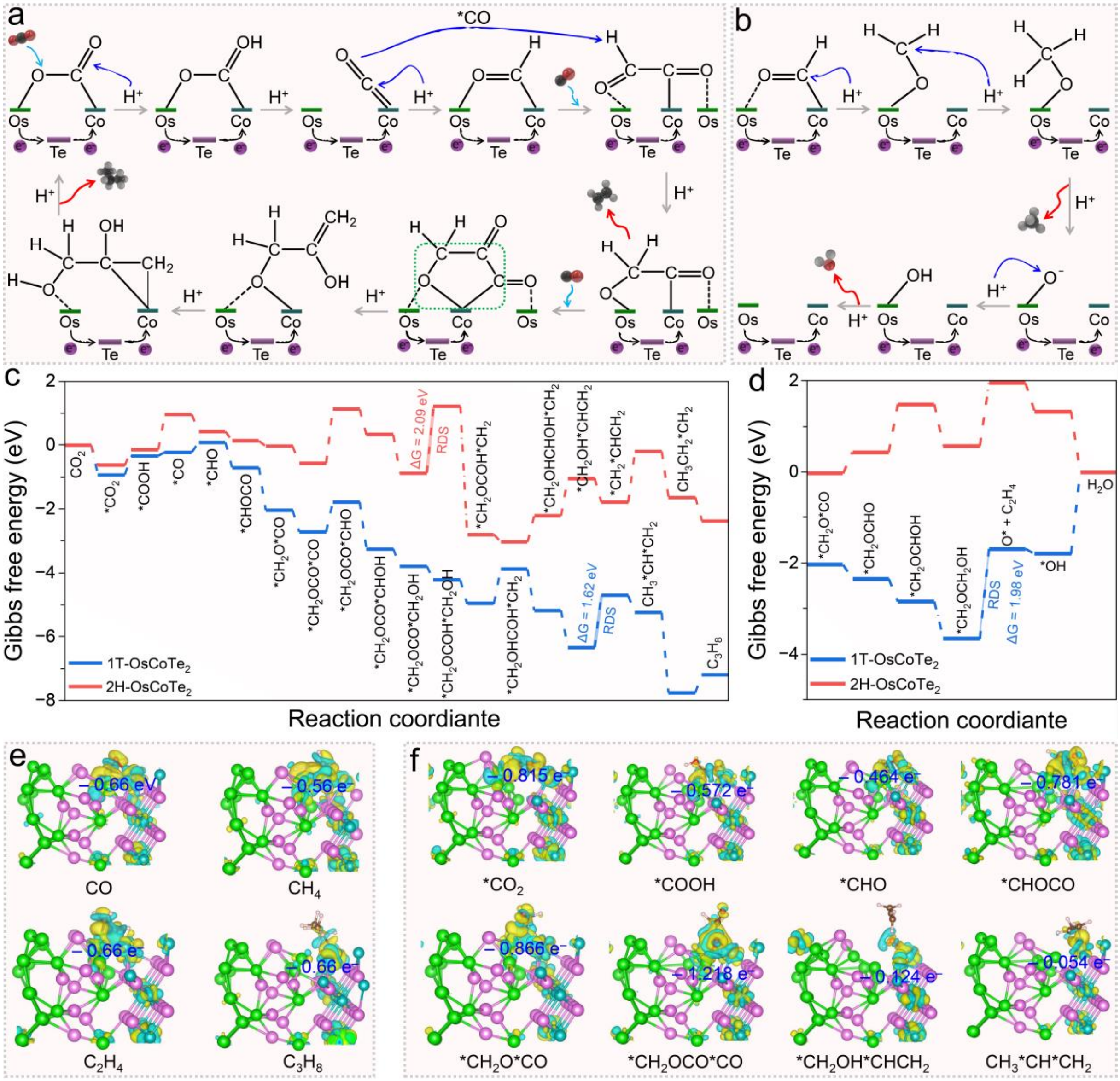

**Fig. 6.** Gibbs free energy diagrams of $CO_2$RR on 1T-$OsCoTe_2$ for (a) $C_2H_4$ and $C_3H_8$, and (b) $CH_4$, respectively. Gibbs free energy for steps in photocatalytic $CO_2$-to-(c) $C_3H_8$ and (d) $C_2H_4$ conversion on 1T-$OsCoTe_2$. (e) Localized charge-density differences of CO, $CH_4$, $C_2H_4$, and $C_3H_8$ on 1T-$OsCoTe_2$. (f) Charge transfer diagrams of 1T-$OsCoTe_2$ during $CO_2$RR.

To further probe the enhanced catalytic performance at the 1T-$OsCoTe_2$, the $CO_2$ activation process is first analyzed. From $CO_2$ adsorption isotherms, 1T-$OsCoTe_2$ has a $CO_2$ adsorption capacity of 49.69 $cm^3\ g^{-1}$ under 1.0 atm at 298 K, which is about 1.31 times larger than that of 2H-$OsCoTe_2$ (37.91 $cm^3\ g^{-1}$, Fig. S18a). $CO_2$ adsorption isotherms reveal that 1T-

$OsCoTe_2$ possesses the highest $CO_2$ uptake capacity, which is the prerequisite step for triggering subsequent $CO_2RR$. Given that 1T-$OsCoTe_2$ has a topological structure, this result suggests that the topological structure plays a crucial role in enhancing $CO_2$ adsorption capacity [13]. Benefiting from the microenvironment constructed via 1T-$CoTe_2$, the optimal 1-$OsCoTe_2$ exhibits a high $CO_2$ adsorption capacity, which is favorable for the $CO_2RR$ to $C_2H_4$ and $C_3H_8$. As revealed by in-situ characterizations, *COOH, *CO, *CHO, and *CHOCO species are the key intermediates for conversion of $CO_2$-to-$C_{2+}$ hydrocarbons, which is closely related to the presence of coupling *CO and Co sites in 1T-$OsCoTe_2$. In order to further elucidate the intrinsic reasons for the excellent catalytic activity of 1T-$OsCoTe_2$, to understand the nature of the photocatalyst active sites, and to analyze the reaction mechanism of coupling *CO in the $C_3H_8$ generation process, DFT calculations were carried out on the optimized 1T-$OsCoTe_2$ and 2H-$OsCoTe_2$ using the computational hydrogen electrode model (Fig. 6e, f, and S19) [34,64]. DFT calculation results show that * CO is difficult to desorb on 1T-$OsCoTe_2$, rather than further hydrogenation or C—C coupling reaction, consistent with the result that CO is the major product on 1T-$OsCoTe_2$ (Fig. 3c, 6c, and S20a). The absorbed $CO_2$ is firstly converted to *CHO through *COOH and *CO on 1T-$OsCoTe_2$ (Fig. S18b–f). Then, *CHO at 1T-$OsCoTe_2$ may couple with the CO diffusing from Co—Te unit to generate *CHOCO with a free energy change of −0.783 eV [55,65]. The $C_1$—$C_2$ coupling reaction (*$CH_2OCO$ + *CO → *$CH_2OCOCO$) is thermodynamically calculated to be a self-reversing proceeding exergonic reaction ($\Delta G$ = −0.677 eV < 0) [66]. It is worth noting that, on 2H-$OsCoTe_2$, $C_1$—$C_1$ coupling is challenging due to the large variation uphill energy changes, resulting in the hydrogenation of *CO to $CH_4$ being preferred (Fig. S20b) [67,68]. Meanwhile, for 1T-$OsCoTe_2$, some of *$C_{2+}$ species will continue hydrogenating through a series of proton-electron steps to form $C_2H_4$. The free energy changes of the rate-determining step (RDS) of $C_3H_8$ formation path (Fig. 6c) and $C_2H_4$ formation path (Fig. 6d) are calculated to be 1.623 and 1.98 e V, respectively, indicating that $C_3H_8$ is easier to generate than $C_2H_4$. This is consistent with our experimental results of outperforming $C_3H_8$ than $C_2H_4$ yields on 1T-$OsCoTe_2$. The downslope energy change of C—C coupling on 1T-$OsCoTe_2$ may be due to the low energy levels of *CHOCO and *$CH_2$OCOCO (Fig. S20c−e). The above analysis shows that both $C_1$—$C_1$ and $C_1$—$C_2$ coupling reactions are favorable on 1T-$OsCoTe_2$, in contrast to 2H-$OsCoTe_2$ and other reported catalysts with challenging C—C coupling.

## 4. Conclusion

A simple solid-phase strategy constructs 1T-$OsCoTe_2$ topological semimetal, meanwhile, the $CO_2$RR properties and the function of Co and Os species in 1T-$OsCoTe_2$ were systematically investigated. In-situ XPS and XAFS investigations reveal that Co and Os atoms coordinated with Te atoms enable an efficient site-to-site electron transfer to ensure the high efficiency of $CO_2$RR. Experimental evidence suggests a high electron-based selectivity of 71.2% for $C_3H_8$ (product-based selectivity of 34.2%) and 84.75% for total $C_{2+}$ hydrocarbons (product-based selectivity of 44.79%), as well as an $IQE_{cr}$ of 54.6% at 380 nm. An in-depth mechanism study illustrates that the synergistic effect of Os—Te—Co electron bridging can produce various surface carbonaceous intermediates, including *$CH_3$ and *CO, and further generate $C_{2+}$ products. In addition, the formation of a five-membered ring between *$CH_2$OCOCO and Co—Te unit of 1T-$OsCoTe_2$, largely alleviating the local electron accumulation and weakens inter- and intramolecular electrostatic repulsion. The electronic and non-trivial TSSs, as well as geometrical effects of 1T-$OsCoTe_2$, may have combined to stabilize the critical *$C_{2+}$ intermediates and lowered their adsorption energy levels, thus facilitating C—C coupling. This study provides insights into the photocatalytic performance of topological semimetallic catalysts and contributes to the design of new topological semimetal with excellent photoactivity. Meanwhile, an in-depth insight of the interfacial active sites at the atomic level is essential to analyze the reaction mechanisms and pathways.

## Supplementary material

Detailed more characteristic of the catalysts and more reduction results.

## Acknowledgments

This work is supported by the Natural Science Foundation of China (11922415, 12274471, 22208331), Guangdong Basic and Applied Basic Research Foundation (2022A1515011168, 2019A1515011718, 2019A1515011337), the Key Research & Development Program of Guangdong Province, China (2019B110209003).

## References

[1] Y. Yang, S. Louisia, S. Yu, J. Jin, I. Roh, C. Chen, M. V. Fonseca Guzman, J. Feijóo, P.-C. Chen, H. Wang, C. J. Pollock, X. Huang, Y.-T. Shao, C. Wang, D. A. Muller, H. D. Abruña, P. Yang, Operando studies reveal active Cu nanograins for $CO_2$ electroreduction, Nature, 614 (2023) 262-269.

[2] C. Heide, Y. Kobayashi, D. R. Baykusheva, D. Jain, J. A. Sobota, M. Hashimoto, P. S. Kirchmann, S. Oh, T. F. Heinz, D. A. Reis, S. Ghimire, Probing topological phase transitions using high-harmonic generation, Nat. Photonics, 16 (2022) 620-624.

[3] G. Li, Y. Xu, Z. Song, Q. Yang, Y. Zhang, J. Liu, U. Gupta, V. Süβ, Y. Sun, P. Sessi, S. S. P. Parkin, B. A. Bernevig, C. Felser, Obstructed Surface States as the Descriptor for Predicting Catalytic Active Sites in Inorganic Crystalline Materials, Adv. Mater., 34 (2022) 2201328.

[4] M. R. Scholz, J. Sánchez-Barriga, D. Marchenko, A. Varykhalov, A. Volykhov, L. V. Yashina, O. Rader, Tolerance of Topological Surface States towards Magnetic Moments: Fe on $Bi_2Se_3$, Phys. Rev. Lett., 108 (2012) 256810.

[5] H. Xie, T. Zhang, R. Xie, Z. Hou, X. Ji, Y. Pang, S. Chen, M.-M. Titirici, H. Weng, G. Chai, Facet Engineering to Regulate Surface States of Topological Crystalline Insulator Bismuth Rhombic Dodecahedrons for Highly Energy Efficient Electrochemical $CO_2$ Reduction, Adv. Mater., 33 (2021) 2008373.

[6] H. Luo, P. Yu, G. Li, K. Yan, Topological quantum materials for energy conversion and storage, Nat. Rev. Phys., 4 (2022) 611-624.

[7] Z. Hu, L. Zhang, A. Chakraborty, G. D'Olimpio, J. Fujii, A. Ge, Y. Zhou, C. Liu, A. Agarwal, I. Vobornik, D. Farias, C.-N. Kuo, C. S. Lue, A. Politano, S.-W. Wang, W. Hu, X. Chen, W. Lu, L. Wang, Terahertz Nonlinear Hall Rectifiers Based on Spin-Polarized Topological Electronic States in 1T-$CoTe_2$, Adv. Mater., 35 (2023) 2209557.

[8] T.-H. Lu, C.-J. Chen, M. Basu, C.-G. Ma, R.-S. Liu, The $CoTe_2$ nanostructure: an efficient and robust catalyst for hydrogen evolution, Chem. Commun., 51 (2015) 17012-17015.

[9] Y. Liu, J. Sun, H. Huang, L. Bai, X. Zhao, B. Qu, L. Xiong, F. Bai, J. Tang, L. Jing, Improving $CO_2$ photoconversion with ionic liquid and Co single atoms, Nat. Commun., 14 (2023) 1457.
[10] Z. Z. Fan, R. C. Luo, Y. X. Zhang, B. Zhang, P. L. Zhai, Y. T. Zhang, C. Wang, J. F. Gao, W. Zhou, L. C. Sun, J. G. Hou, Oxygen-Bridged Indium-Nickel Atomic Pair as Dual-Metal Active Sites Enabling Synergistic Electrocatalytic $CO_2$ Reduction, Angew. Chem. Int. Ed., (2023).
[11] H.-J. Liu, S. Zhang, Y.-M. Chai, B. Dong, Ligand Modulation of Active Sites to Promote Cobalt-Doped 1T-$MoS_2$ Electrocatalytic Hydrogen Evolution in Alkaline Media, Angew. Chem. Int. Ed., 62 (2023) e202313845.
[12] J. Wang, G. Wang, J. Zhang, Y. Wang, H. Wu, X. Zheng, J. Ding, X. Han, Y. Deng, W. Hu, Inversely Tuning the $CO_2$ Electroreduction and Hydrogen Evolution Activity on Metal Oxide via Heteroatom Doping, Angew. Chem. Int. Ed., 60 (2021) 7602-7606.
[13] Q. Zhang, S. Gao, Y. Guo, H. Wang, J. Wei, X. Su, H. Zhang, Z. Liu, J. Wang, Designing covalent organic frameworks with Co-$O_4$ atomic sites for efficient $CO_2$ photoreduction, Nat. Commun., 14 (2023) 1147.
[14] G. Kresse, J. Hafner, Ab initio molecular dynamics for open-shell transition metals, Phys. Rev. B, 48 (1993) 13115-13118.
[15] W. Kohn, L. J. Sham, Self-Consistent Equations Including Exchange and Correlation Effects, Physical Review, 140 (1965) A1133-A1138.
[16] R. J. Protheroe, T. Stanev, Limits on Models of the Ultrahigh Energy Cosmic Rays Based on Topological Defects [Phys. Rev. Lett. 77, 3708 (1996)], Phys. Rev. Lett., 78 (1997) 3420-3420.
[17] S. Grimme, J. Antony, S. Ehrlich, H. Krieg, A consistent and accurate ab initio parametrization of density functional dispersion correction (DFT-D) for the 94 elements H-Pu, J. Chem. Phys., 132 (2010) 154104.
[18] S. Grimme, S. Ehrlich, L. Goerigk, Effect of the damping function in dispersion corrected density functional theory, J. Comput. Chem., 32 (2011) 1456-1465.
[19] V. I. Anisimov, J. Zaanen, O. K. Andersen, Band theory and Mott insulators: Hubbard U instead of Stoner I, Phys. Rev. B, 44 (1991) 943-954.
[20] F. Li, X. Yue, Y. Liao, L. Qiao, K. Lv, Q. Xiang, Understanding the unique S-scheme charge migration in triazine/heptazine crystalline carbon nitride homojunction, Nat. Commun., 14 (2023) 3901.

[21] K. Wang, Z. Huang, J. Wang, Synthesis of hierarchical tandem double Z-scheme heterojunctions for robust photocatalytic $H_2$ generation, Chem. Eng. J., 430 (2022) 132727.
[22] M. Samanta, H. Tan, S. Laha, H. A. Vignolo-González, L. Grunenberg, S. Bette, V. Duppel, P. Schützendübe, A. Gouder, B. Yan, B.V. Lotsch, The Weyl Semimetals $MIrTe_4$ (M = Nb, Ta) as Efficient Catalysts for Dye-Sensitized Hydrogen Evolution, Adv. Energy Mater., 13 (2023) 2300503.
[23] X. Chia, Z. Sofer, J. Luxa, M. Pumera, Unconventionally Layered $CoTe_2$ and $NiTe_2$ as Electrocatalysts for Hydrogen Evolution, Chem. Eur. J., 23 (2017) 11719-11726.
[24] J. Liu, Y. Shen, X. Gao, L. Lv, Y. Ma, S. Wu, X. Wang, Z. Zhou, $GeN_3$ monolayer: A promising 2D high-efficiency photo- hydrolytic catalyst with High carrier mobility transport anisotropy, Appl. Catal. B: Environ., 279 (2020) 119368.
[25] M. Li, H. Zhu, Q. Yuan, T. Li, M. Wang, P. Zhang, Y. Zhao, D. Qin, W. Guo, B. Liu, X. Yang, Y. Liu, Y. Pan, Proximity Electronic Effect of Ni/Co Diatomic Sites for Synergistic Promotion of Electrocatalytic Oxygen Reduction and Hydrogen Evolution, Adv. Funct. Mater., 33 (2023) 2210867.
[26] J. Zhu, F. Xia, Y. Guo, R. Lu, L. Gong, D. Chen, P. Wang, L. Chen, J. Yu, J. Wu, S. Mu, Electron Accumulation Effect over Osmium/Erlichmanite Heterointerfaces for Intensified pH-Universal Hydrogen Evolution, ACS Catal., 12 (2022) 13312-13320.
[27] R. Huang, Y. Wen, P. Miao, W. Shi, W. Niu, K. Sun, Y. Li, Y. Ji, B. Zhang, Constructing the oxygen diffusion paths for promoting the stability of acidic water oxidation catalysts, Chem Catalysis, 3 (2023) 100667.
[28] Y. Jiang, Y.-P. Deng, R. Liang, N. Chen, G. King, A. Yu, Z. Chen, Linker-Compensated Metal–Organic Framework with Electron Delocalized Metal Sites for Bifunctional Oxygen Electrocatalysis, J. Am. Chem. Soc., 144 (2022) 4783-4791.
[29] N. Wang, S. Ning, X. Yu, D. Chen, Z. Li, J. Xu, H. Meng, D. Zhao, L. Li, Q. Liu, B. Lu, S. Chen, Graphene composites with Ru-$RuO_2$ heterostructures: Highly efficient Mott–Schottky-type electrocatalysts for pH-universal water splitting and flexible zinc–air batteries, Appl. Catal. B: Environ., 302 (2022) 120838.
[30] L. Shi, X. Ren, Q. Wang, Y. Li, F. Ichihara, H. Zhang, Y. Izumi, L. Ren, W. Zhou, Y. Yang, J. Ye, Stabilizing Atomically Dispersed Catalytic Sites on Tellurium Nanosheets with Strong Metal–Support Interaction Boosts Photocatalysis, Small, 16 (2020) 2002356.
[31] Z. Jiang, X. Liu, X.-Z. Liu, S. Huang, Y. Liu, Z.-C. Yao, Y. Zhang, Q.-H. Zhang, L. Gu, L.-R. Zheng, L. Li, J. Zhang, Y. Fan, T. Tang, Z. Zhuang, J.-S. Hu, Interfacial assembly of

binary atomic metal-Nx sites for high-performance energy devices, Nat. Commun., 14 (2023) 1822.
[32] X.-Y. Dong, Y.-N. Si, Q.-Y. Wang, S. Wang, S.-Q. Zang, Integrating Single Atoms with Different Microenvironments into One Porous Organic Polymer for Efficient Photocatalytic $CO_2$ Reduction, Adv. Mater., 33 (2021) 2101568.
[33] S. Yang, W. Hu, X. Zhang, P. He, B. Pattengale, C. Liu, M. Cendejas, I. Hermans, X. Zhang, J. Zhang, J. Huang, 2D Covalent Organic Frameworks as Intrinsic Photocatalysts for Visible Light-Driven $CO_2$ Reduction, J. Am. Chem. Soc., 140 (2018) 14614-14618.
[34] Y. Shen, C. Ren, L. Zheng, X. Xu, R. Long, W. Zhang, Y. Yang, Y. Zhang, Y. Yao, H. Chi, J. Wang, Q. Shen, Y. Xiong, Z. Zou, Y. Zhou, Room-temperature photosynthesis of propane from CO2 with Cu single atoms on vacancy-rich $TiO_2$, Nat. Commun., 14 (2023) 1117.
[35] C. Ding, X. Lu, B. Tao, L. Yang, X. Xu, L. Tang, H. Chi, Y. Yang, D.M. Meira, L. Wang, X. Zhu, S. Li, Y. Zhou, Z. Zou, Interlayer Spacing Regulation by Single-Atom Indium$^{\delta+}$–$N_4$ on Carbon Nitride for Boosting $CO_2$/CO Photo-Conversion, Adv. Funct. Mater., n/a (2023) 2302824.
[36] K. Wang, Z. Hu, P. Yu, A.M. Balu, K. Li, L. Li, L. Zeng, C. Zhang, R. Luque, K. Yan, H. Luo, Understanding Bridging Sites and Accelerating Quantum Efficiency for Photocatalytic CO2 Reduction, Nano-Micro Lett., 16 (2023) 5.
[37] J. Li, H. Huang, W. Xue, K. Sun, X. Song, C. Wu, L. Nie, Y. Li, C. Liu, Y. Pan, H.-L. Jiang, D. Mei, C. Zhong, Self-adaptive dual-metal-site pairs in metal-organic frameworks for selective $CO_2$ photoreduction to $CH_4$, Nat. Catal., 4 (2021) 719-729.
[38] H. Jiang, K.-i. Katsumata, J. Hong, A. Yamaguchi, K. Nakata, C. Terashima, N. Matsushita, M. Miyauchi, A. Fujishima, Photocatalytic reduction of $CO_2$ on $Cu_2O$-loaded Zn-Cr layered double hydroxides, Appl. Catal. B: Environ., 224 (2018) 783-790.
[39] Q. Zhi, J. Zhou, W. Liu, L. Gong, W. Liu, H. Liu, K. Wang, J. Jiang, Covalent Microporous Polymer Nanosheets for Efficient Photocatalytic $CO_2$ Conversion with $H_2O$, Small, 18 (2022) 2201314.
[40] C. Hu, Y. Zhang, A. Hu, Y. Wang, X. Wei, K. Shen, L. Chen, Y. Li, Near- and Long-Range Electronic Modulation of Single Metal Sites to Boost $CO_2$ Electrocatalytic Reduction, Adv. Mater., 35 (2023) 2209298.
[41] W. Gao, X. Zhao, T. Zhang, X. Yu, Y. Ma, E.C. dos Santos, J. White, H. Liu, Y. Sang, Construction of diluted magnetic semiconductor to endow nonmagnetic semiconductor with spin-regulated photocatalytic performance, Nano Energy, 110 (2023) 108381.

[42] K. Wang, Z. Huang, X. Jin, D. Zhang, J. Wang, MOF–derived hollow porous $ZnFe_2O_4$/AgCl/Ag/C nanotubes with magnetic–dielectric synergy as high–performance photocatalysts for hydrogen evolution reaction, Chem. Eng. J., 422 (2021) 130140.
[43] K. Wang, S. Zhan, H. Sun, D. Zhang, J. Wang, Hollow porous core–shell $ZnFe_2O_4$/AgCl nanocubes coated with EDTA and Ag nanoparticles for enhanced photocatalytic performances of visible–light–driven, Chem. Eng. J., 400 (2020) 125908.
[44] R. Li, T. Takata, B. Zhang, C. Feng, Q. Wu, C. Cui, Z. Zhang, K. Domen, Y. Li, Criteria for Efficient Photocatalytic Water Splitting Revealed by Studying Carrier Dynamics in a Model Al-doped $SrTiO_3$ Photocatalyst, Angew. Chem. Int. Ed., 62 (2023) e202313537.
[45] H. Yu, Y. Wang, X. Zou, H. Han, H. K. Kim, Z. Yao, Z. Wang, Y. Li, H. M. Ng, W. Zhou, J. Zhang, S. Chen, X. Lu, K. S. Wong, Z. Zhu, H. Yan, H. Hu, Effects of Halogenation of Small-Molecule and Polymeric Acceptors for Efficient Organic Solar Cells, Adv. Funct. Mater., 33 (2023) 2300712.
[46] A. M. El-Zohry, B. Turedi, A. Alsalloum, P. Maity, O.M. Bakr, B. S. Ooi, O. F. Mohammed, Ultrafast transient infrared spectroscopy for probing trapping states in hybrid perovskite films, Commun. Chem., 5 (2022) 67.
[47] J. Ran, H. Zhang, S. Fu, M. Jaroniec, J. Shan, B. Xia, Y. Qu, J. Qu, S. Chen, L. Song, J. M. Cairney, L. Jing, S.-Z. Qiao, $NiPS_3$ ultrathin nanosheets as versatile platform advancing highly active photocatalytic $H_2$ production, Nat. Commun., 13 (2022) 4600.
[48] W. Huang, C. Su, C. Zhu, T. Bo, S. Zuo, W. Zhou, Y. Ren, Y. Zhang, J. Zhang, M. Rueping, Isolated electron trap-induced charge accumulation for efficient photocatalytic hydrogen production, Angewandte Chemie, 135 (2023) e202304634.
[49] J. Kosco, S. Gonzalez-Carrero, C. T. Howells, T. Fei, Y. Dong, R. Sougrat, G. T. Harrison, Y. Firdaus, R. Sheelamanthula, B. Purushothaman, F. Moruzzi, W. Xu, L. Zhao, A. Basu, S. De Wolf, T. D. Anthopoulos, J. R. Durrant, I. McCulloch, Generation of long-lived charges in organic semiconductor heterojunction nanoparticles for efficient photocatalytic hydrogen evolution, Nat. Energy, 7 (2022) 340-351.
[50] H. Yu, J. Li, Y. Zhang, S. Yang, K. Han, F. Dong, T. Ma, H. Huang, Three-in-One Oxygen Vacancies: Whole Visible-Spectrum Absorption, Efficient Charge Separation, and Surface Site Activation for Robust $CO_2$ Photoreduction, Angew. Chem. Int. Ed., 58 (2019) 3880-3884.
[51] Q. Liu, C. Chen, K. Yuan, C. D. Sewell, Z. Zhang, X. Fang, Z. Lin, Robust route to highly porous graphitic carbon nitride microtubes with preferred adsorption ability via rational design of one-dimension supramolecular precursors for efficient photocatalytic $CO_2$ conversion, Nano Energy, 77 (2020) 105104.

[52] C. Ban, Y. Duan, Y. Wang, J. Ma, K. Wang, J. Meng, X. Liu, C. Wang, X. Han, G. Cao, L. Gan, X. Zhou, Isotype Heterojunction-Boosted $CO_2$ Photoreduction to CO, Nano-Micro Lett., 14 (2022) 74.
[53] S. Bai, W. Jing, G. He, C. Liao, F. Wang, Y. Liu, L. Guo, Near-Infrared-Responsive Photocatalytic $CO_2$ Conversion via In Situ Generated $Co_3O_4/Cu_2O$, ACS Nano, 17 (2023) 10976-10986.
[54] W. Zhang, D. Xi, Y. Chen, A. Chen, Y. Jiang, H. Liu, Z. Zhou, H. Zhang, Z. Liu, R. Long, Y. Xiong, Light-driven flow synthesis of acetic acid from methane with chemical looping, Nat. Commun., 14 (2023) 3047.
[55] X.-F. Qiu, H.-L. Zhu, J.-R. Huang, P.-Q. Liao, X.-M. Chen, Highly Selective $CO_2$ Electroreduction to $C_2H_4$ Using a Metal–Organic Framework with Dual Active Sites, J. Am. Chem. Soc., 143 (2021) 7242-7246.
[56] J. Liang, H. Zhang, Q. Song, Z. Liu, J. Xia, B. Yan, X. Meng, Z. Jiang, X. W. Lou, C.-S. Lee, Modulating Charge Separation of Oxygen-Doped Boron Nitride with Isolated Co Atoms for Enhancing $CO_2$-to-CO Photoreduction, Adv. Mater., 36 (2024) 2303287.
[57] S. Zhang, C. Huang, Z. Shao, H. Zhou, J. Chen, L. Li, J. Lu, X. Liu, H. Luo, L. Xia, H. Wang, Y. Sun, Revealing and Regulating the Complex Reaction Mechanism of $CO_2$ Hydrogenation to Higher Alcohols on Multifunctional Tandem Catalysts, ACS Catal., 13 (2023) 3055-3065.
[58] H. Xu, D. Rebollar, H. He, L. Chong, Y. Liu, C. Liu, C.-J. Sun, T. Li, J. V. Muntean, R. E. Winans, D.-J. Liu, T. Xu, Highly selective electrocatalytic $CO_2$ reduction to ethanol by metallic clusters dynamically formed from atomically dispersed copper, Nat. Energy, 5 (2020) 623-632.
[59] J. Chen, J. Zhong, Y. Wu, W. Hu, P. Qu, X. Xiao, G. Zhang, X. Liu, Y. Jiao, L. Zhong, Y. Chen, Particle Size Effects in Stoichiometric Methane Combustion: Structure–Activity Relationship of Pd Catalyst Supported on Gamma-Alumina, ACS Catal., 10 (2020) 10339-10349.
[60] T. Cheng, H. Xiao, W. A. Goddard, III, Reaction Mechanisms for the Electrochemical Reduction of $CO_2$ to CO and Formate on the Cu(100) Surface at 298 K from Quantum Mechanics Free Energy Calculations with Explicit Water, J. Am. Chem. Soc., 138 (2016) 13802-13805.
[61] J. Zhu, W. Shao, X. Li, X. Jiao, J. Zhu, Y. Sun, Y. Xie, Asymmetric Triple-Atom Sites Confined in Ternary Oxide Enabling Selective $CO_2$ Photothermal Reduction to Acetate, J. Am. Chem. Soc., 143 (2021) 18233-18241.

[62] W. Wang, C. Deng, S. Xie, Y. Li, W. Zhang, H. Sheng, C. Chen, J. Zhao, Photocatalytic C–C Coupling from Carbon Dioxide Reduction on Copper Oxide with Mixed-Valence Copper(I)/Copper(II), J. Am. Chem. Soc., 143 (2021) 2984-2993.
[63] M. Esmaeilirad, Z. Jiang, A. M. Harzandi, A. Kondori, M. Tamadoni Saray, C. U. Segre, R. Shahbazian-Yassar, A. M. Rappe, M. Asadi, Imidazolium-functionalized $Mo_3P$ nanoparticles with an ionomer coating for electrocatalytic reduction of $CO_2$ to propane, Nat. Energy, 8 (2023) 891-900.
[64] X. Liu, J. Xiao, H. Peng, X. Hong, K. Chan, J. K. Nørskov, Understanding trends in electrochemical carbon dioxide reduction rates, Nat. Commun., 8 (2017) 15438.
[65] A. J. Garza, A.T. Bell, M. Head-Gordon, Mechanism of $CO_2$ Reduction at Copper Surfaces: Pathways to $C_2$ Products, ACS Catal., 8 (2018) 1490-1499.
[66] C. Du, J. P. Mills, A. G. Yohannes, W. Wei, L. Wang, S. Lu, J.-X. Lian, M. Wang, T. Guo, X. Wang, H. Zhou, C.-J. Sun, J.Z. Wen, B. Kendall, M. Couillard, H. Guo, Z. Tan, S. Siahrostami, Y.A. Wu, Cascade electrocatalysis via AgCu single-atom alloy and Ag nanoparticles in $CO_2$ electroreduction toward multicarbon products, Nat. Commun., 14 (2023) 6142.
[67] X. Li, Y. Sun, J. Xu, Y. Shao, J. Wu, X. Xu, Y. Pan, H. Ju, J. Zhu, Y. Xie, Selective visible-light-driven photocatalytic $CO_2$ reduction to $CH_4$ mediated by atomically thin $CuIn_5S_8$ layers, Nat. Energy, 4 (2019) 690-699.
[68] Z.-W. Wang, Q. Wan, Y.-Z. Shi, H. Wang, Y.-Y. Kang, S.-Y. Zhu, S. Lin, L. Wu, Selective photocatalytic reduction $CO_2$ to $CH_4$ on ultrathin $TiO_2$ nanosheet via coordination activation, Appl. Catal. B: Environ., 288 (2021) 120000.

# Supporting Information

## Non-trivial Topological Surface States Regulation of 1T-$OsCoTe_2$ Enables Selective C—C Coupling for Highly Efficient Photochemical $CO_2$ Reduction Toward $C_{2+}$ hydrocarbons

Kangwang Wang[1], Mingjie Wu[2], Peifeng Yu[1], Hector F. Garces[3], Ying Liang[4], Longfu Li[1], Lingyong Zeng[1], Kuan Li[1], Chao Zhang[1], Kai Yan[5,*], Huixia Luo[1,*]

1 School of Materials Science and Engineering, State Key Laboratory of Optoelectronic Materials and Technologies, Guangdong Provincial Key Laboratory of Magnetoelectric Physics and Devices, Key Lab of Polymer Composite & Functional Materials, Sun Yat-sen University, Guangzhou 510275, China

2 State Key Laboratory of New Textile Materials and Advanced Processing Technologies, Wuhan Textile University, Wuhan 430200, China

3 School of Engineering, Brown University, 182 Hope Street, Providence, USA.

4 The Basic Course Department, Guangzhou Maritime University, Guangzhou 510800, China

5 School of Environmental Science and Engineering, Sun Yat-sen University, Guangzhou 510275, China

* Corresponding author. Email: yank9@mail.sysu.edu.cn (K. Yan); luohx7@mail.sysu.edu.cn (H. X. Luo)

## Section I: Characterization

**X-ray absorption fine structure spectroscopy (XAFS):** The Co *K*-edge and Os $L_3$-edge XAS measurements were conducted at the BL14W1 station within the Shanghai Synchrotron Radiation Facility (SSRF). EXAFS fitting was applied through Athena and Artemis software [1]. WT spectra was also employed using the software package developed by Funke and Chukalina using Morlet wavelet with $\kappa = 10$ and $\sigma = 1$. The data in the XANES region of the absorption coefficient were examined via applying the same procedure for pre-edge line fitting, post edge curve fitting, and edge step normalization to all data.

**In-situ DRIFTS:** In-situ DRIFTS spectra was recorded to detect the intermediates during $CO_2$RR process. The reaction chamber was first bubbled with Ar, and then filled with $CO_2$. The data were collected periodically under the irradiation of a 300 W Xe lamp with a UVCUT-420 nm filter. In-situ DRIFTS spectra characterization was operated on a Nicolet iS50 FTIR spectrometer from 2000 ~ 1000 $cm^{-1}$.

**In-situ NAP-XPS:** The in-situ NAP-XPS measurements were obtained on the beamline BL02B1 of SSRF equipped with a 300 W Xe lamp as the illumination source.

**Radical intermediate characterizations:** Using 5,5-dimethyl-1-pyrroline N-oxide (DMPO) as the radical trapping agent, in-situ ESR spectra of $CO_2$+$H_2O$ mixture under light illumination over 1T-$OsCoTe_2$ and 2H-$OsCoTe_2$ at 298 K. Photocatalyst amount: 3 mg.

## Section II: Figures and Tables

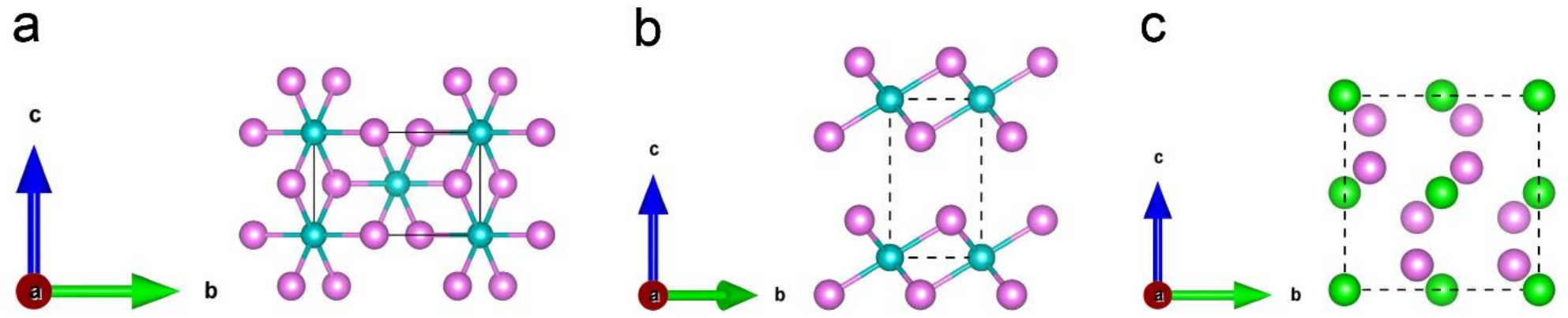

**Fig. S1.** Crystal structures of (a) 2H-$CoTe_2$, (b) 1T-$CoTe_2$, and (c) 2H-$OsTe_2$, respectively.

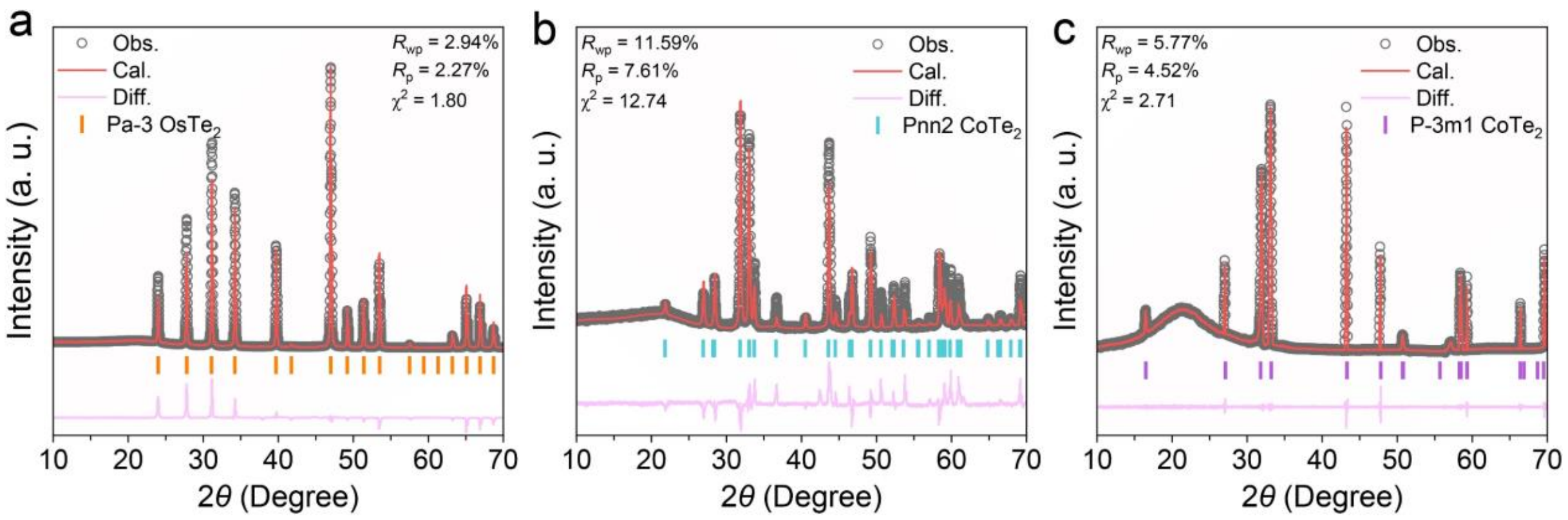

**Fig. S2.** (a) 2H-$OsTe_2$, (b) 2H-$CoTe_2$, and (c) 1T-$CoTe_2$, respectively.

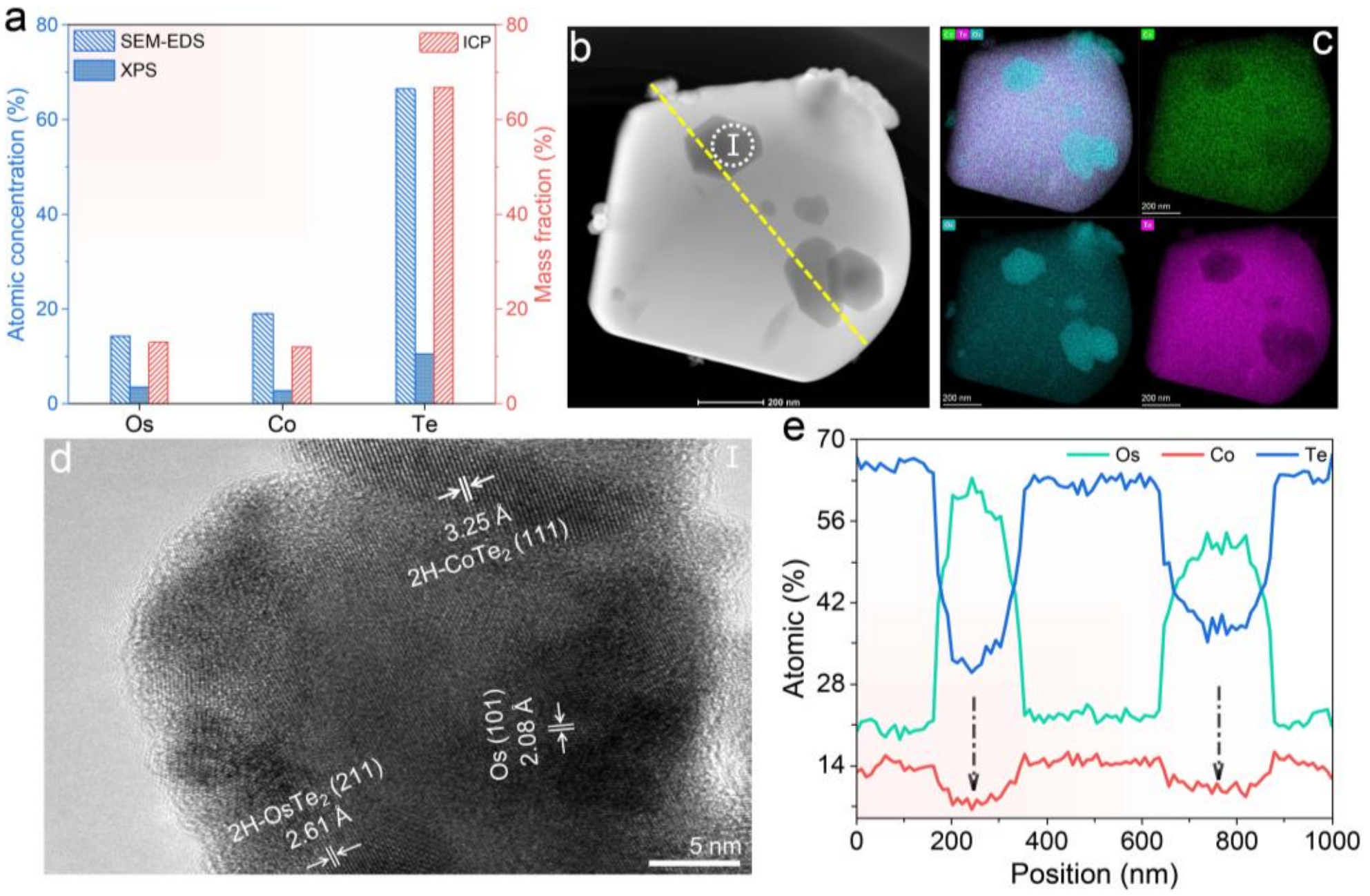

**Fig. S3.** (a) Quantitative data for the main metallic elements in 1T-$OsCoTe_2$, as determined with XPS and inductively coupled plasma-mass spectrometry. (b) HAADF-STEM, (c) the corresponding STEM-EDX mapping, and (d) HRTEM images of 2H-$OsCoTe_2$. (e) Line-scanning analysis along the yellow and blue arrows in (b), where the arrows represent Co scarcity.

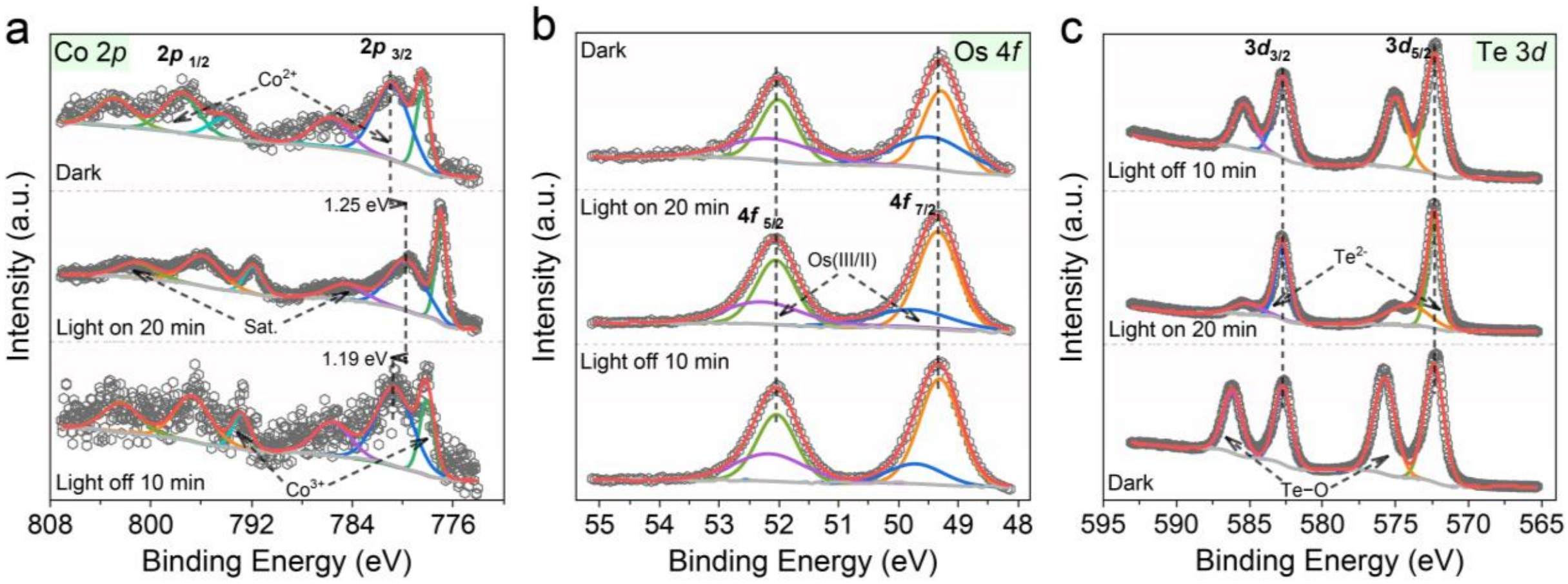

**Fig. S4.** In-situ XPS (a) Co 2*p*, (b) Os 4*f*, and (c) Te 3*d* spectra of 2H-$OsCoTe_2$.

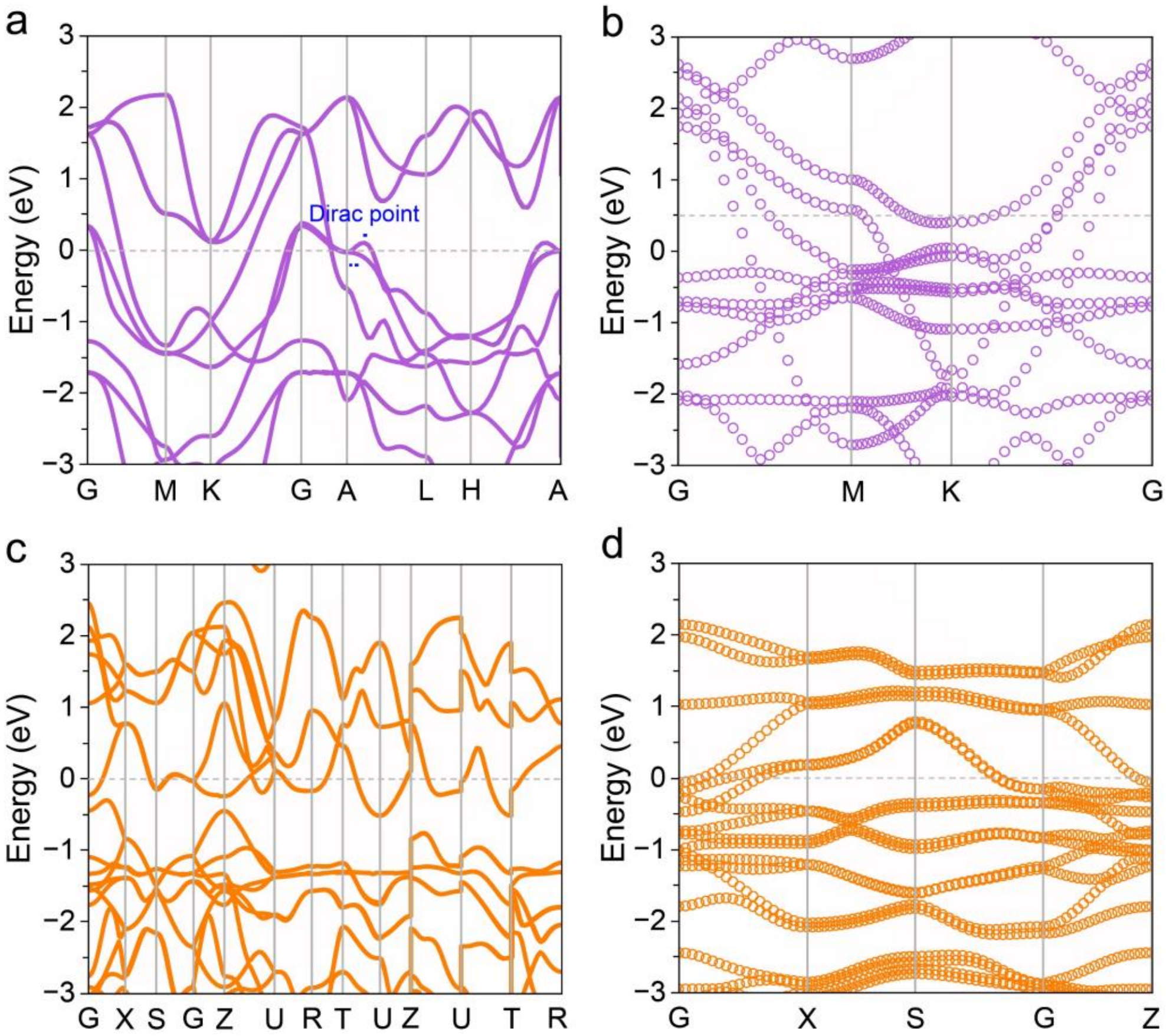

**Fig. S5.** Band structure of (a) 1T-$CoTe_2$ and (c) 2H-$OsTe_2$ without SOC. The surface energy band of (b)1T-$CoTe_2$ and (d) 2H-$OsTe_2$ (001) surface.

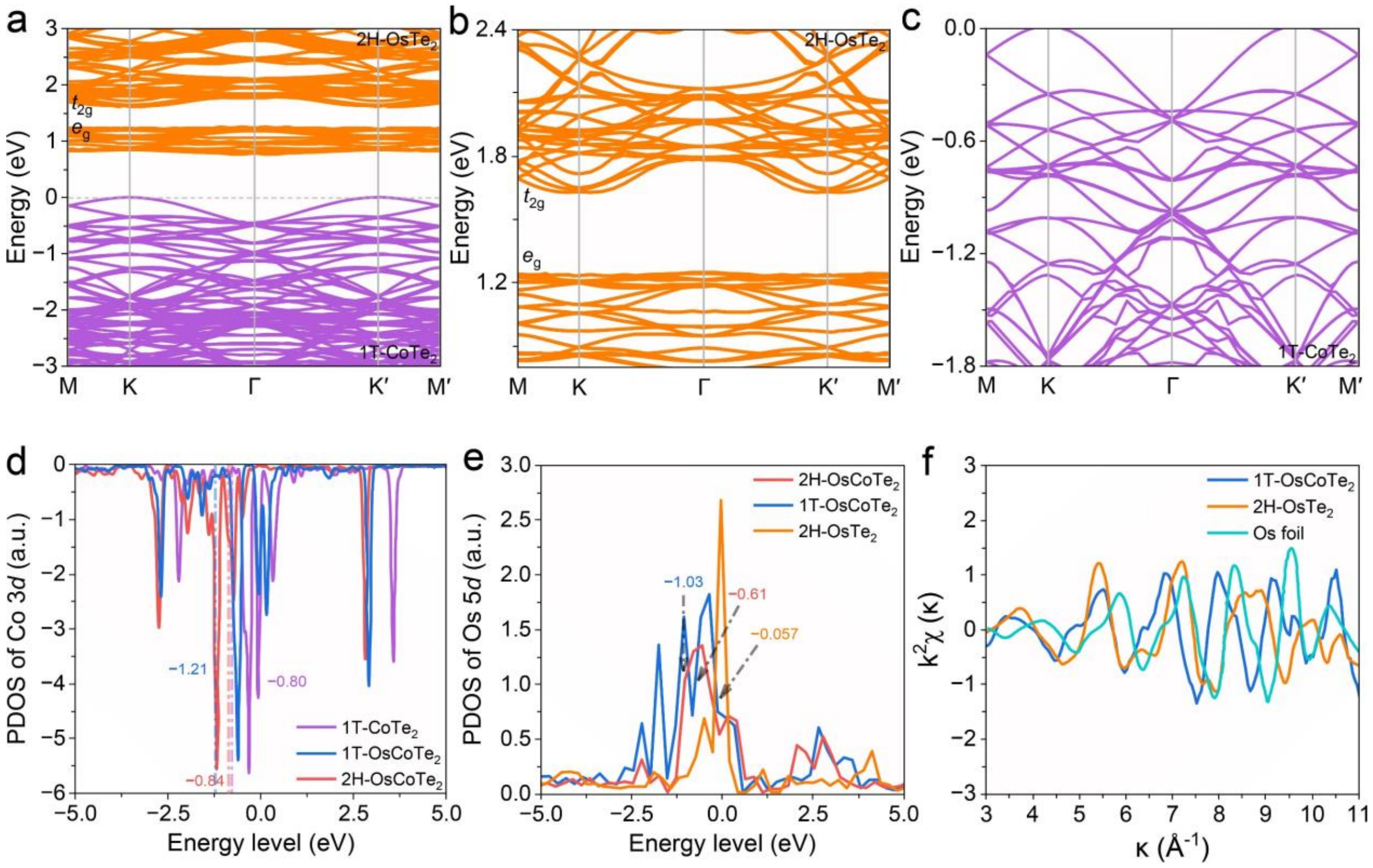

**Fig. S6.** (a–c) The surface energy band of 1T-$OsCoTe_2$ (001) surface. PDOS and corresponding calculated *d*-band centers of (d) Co 3*d* and (e) Os 5*d* for 1T-$OsCoTe_2$, 2H-$OsCoTe_2$, 1T-$CoTe_2$, and 2H-$OsTe_2$. (f) Os *K*-space oscillation, weighted by $k^2$.

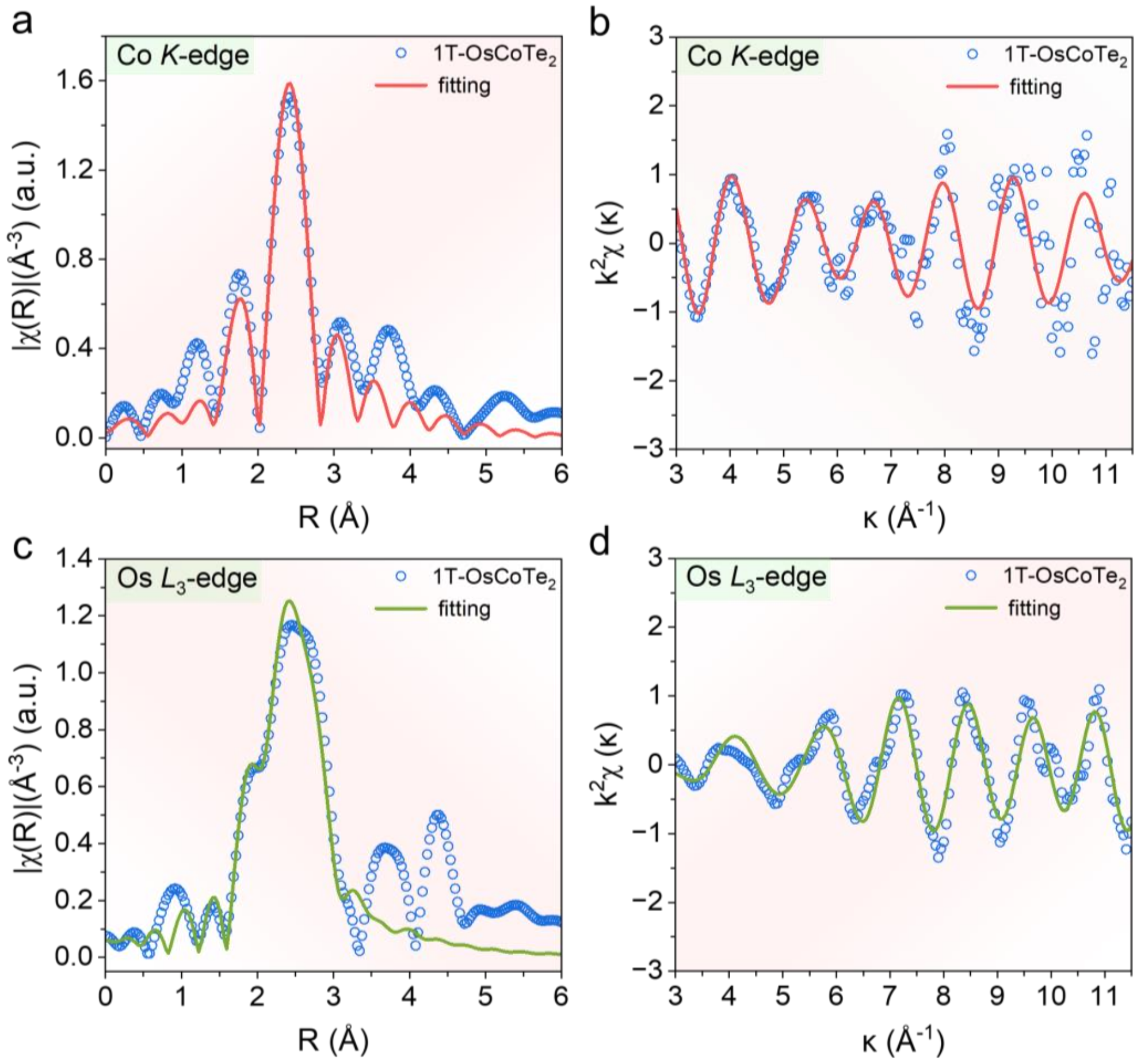

**Fig. S7.** Fitting results of EXAFS spectra in (a,c) R-space and (b,d) k-space for 1T-$OsCoTe_2$.

**Table S1** Fitting parameters for Co *K*-edge and Os $L_3$-edge EXAFS signals for 1T-$OsCoTe_2$.

| Sample | Paths | CN | R (Å) | $\sigma^2$ ($10^{-2}$ Å$^2$) | $\Delta E_0$ | r-factor (%) |
|---|---|---|---|---|---|---|
| Co foil | **Co**–Co | 12 | 2.25 | 0.6 | 6.1 | 0.1 |
| 1T-$OsCoTe_2$ | **Co**–O | 2.7 | 2.49 | 0.3 | 18.3 | 0.8 |
| | **Co**–Co | 2.4 | 2.55 | 0.3 | 18.3 | |
| | **Co**–Os | 1.7 | 2.62 | 0.3 | 18.3 | |
| Os foil | **Os**–Os | 12 | 2.69 | 0.3 | 7.0 | 0.2 |
| 1T-$OsCoTe_2$ | **Os**–Co | 0.9 | 2.59 | 0.1 | 1.8 | 1.2 |
| | **Os**–Os | 5.5 | 2.68 | 0.1 | 1.8 | |

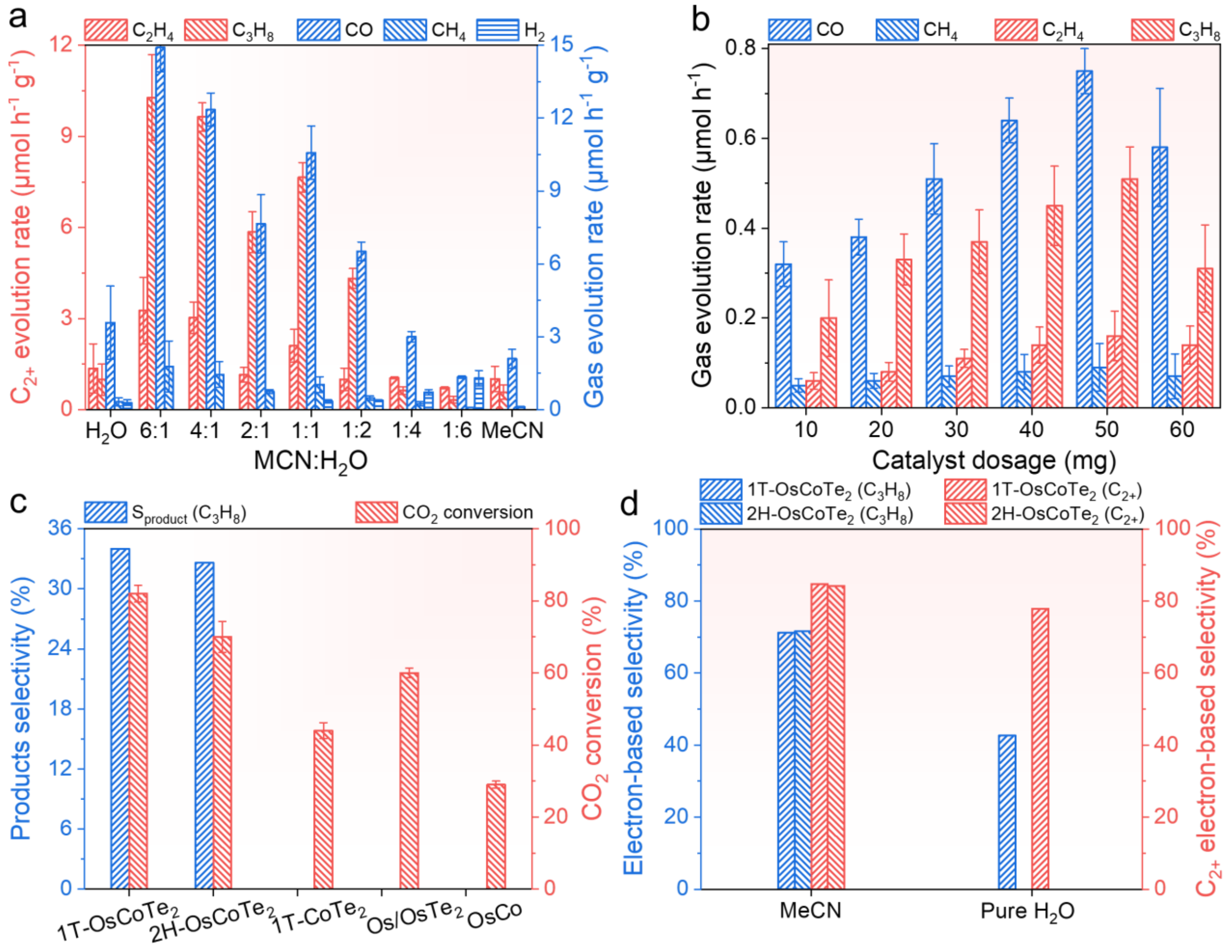

**Fig. S8.** (a) Effect of the MeCN/$H_2O$ volume ratio on the photocatalytic $CO_2$RR. (b) Dependence of catalyst mass-loading on photocatalytic gas production within 1 h. (c) $CO_2$ conversion and product selectivity of 1T-$OsCoTe_2$ in MeCN aqueous solution. (d) Comparison for selectivity of total $C_{2+}$ products and $C_3H_8$ over 1T-$OsCoTe_2$ and 2H-$OsCoTe_2$ in MeCN aqueous solution and pure $H_2O$.

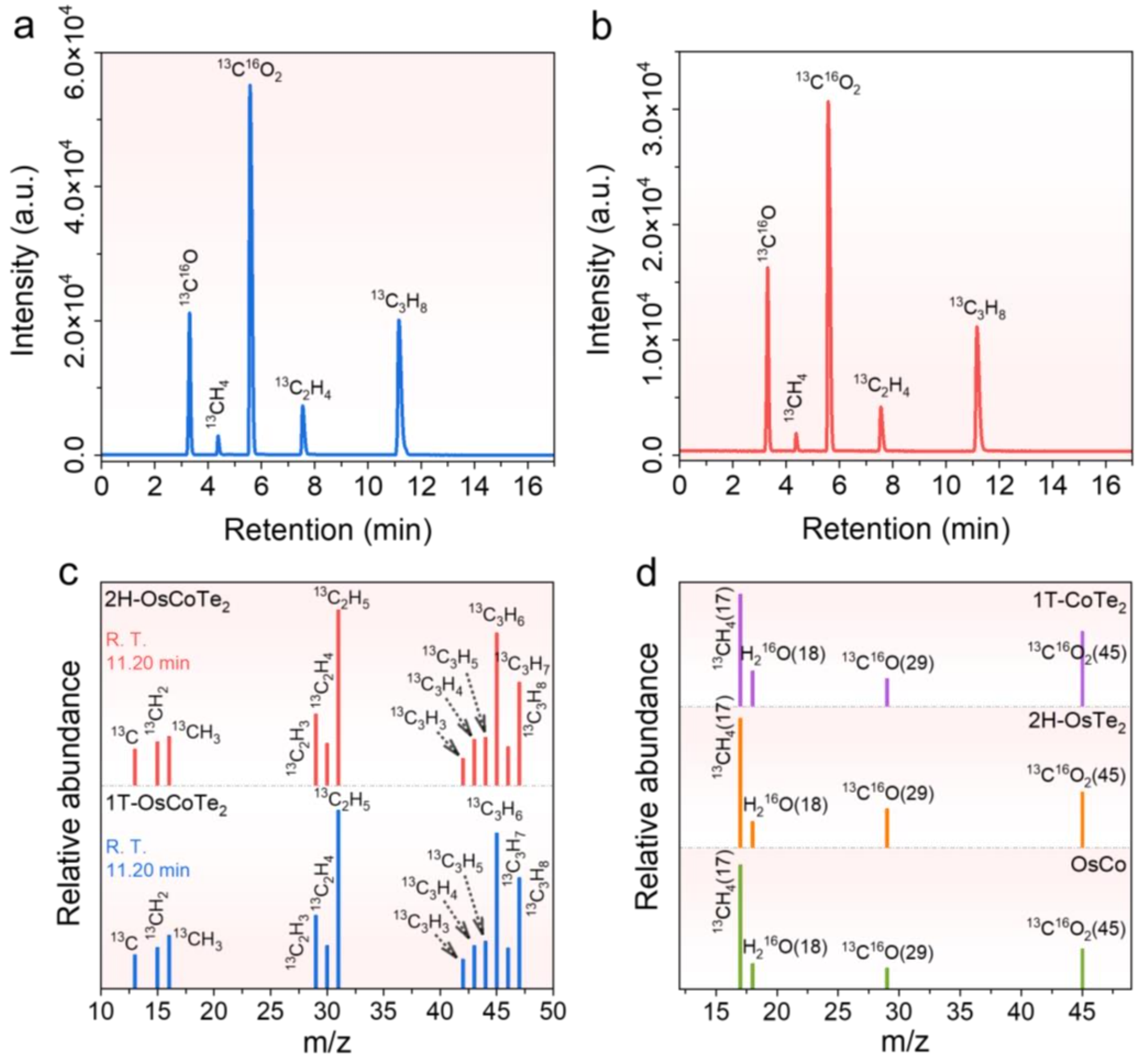

**Fig. S9.** GC spectra for $^{13}CH_4$, $^{13}CO_2$, $^{13}C_2H_4$, and $^{13}C_3H_8$ using HP-PLOT/Q at 350 K. MS spectra of $^{13}CO$, $^{13}CH_4$, $^{13}C_2H_4$, and $^{13}C_3H_8$ produced from the photocatalytic reduction of $^{13}CO_2$ for different photocatalytic system.

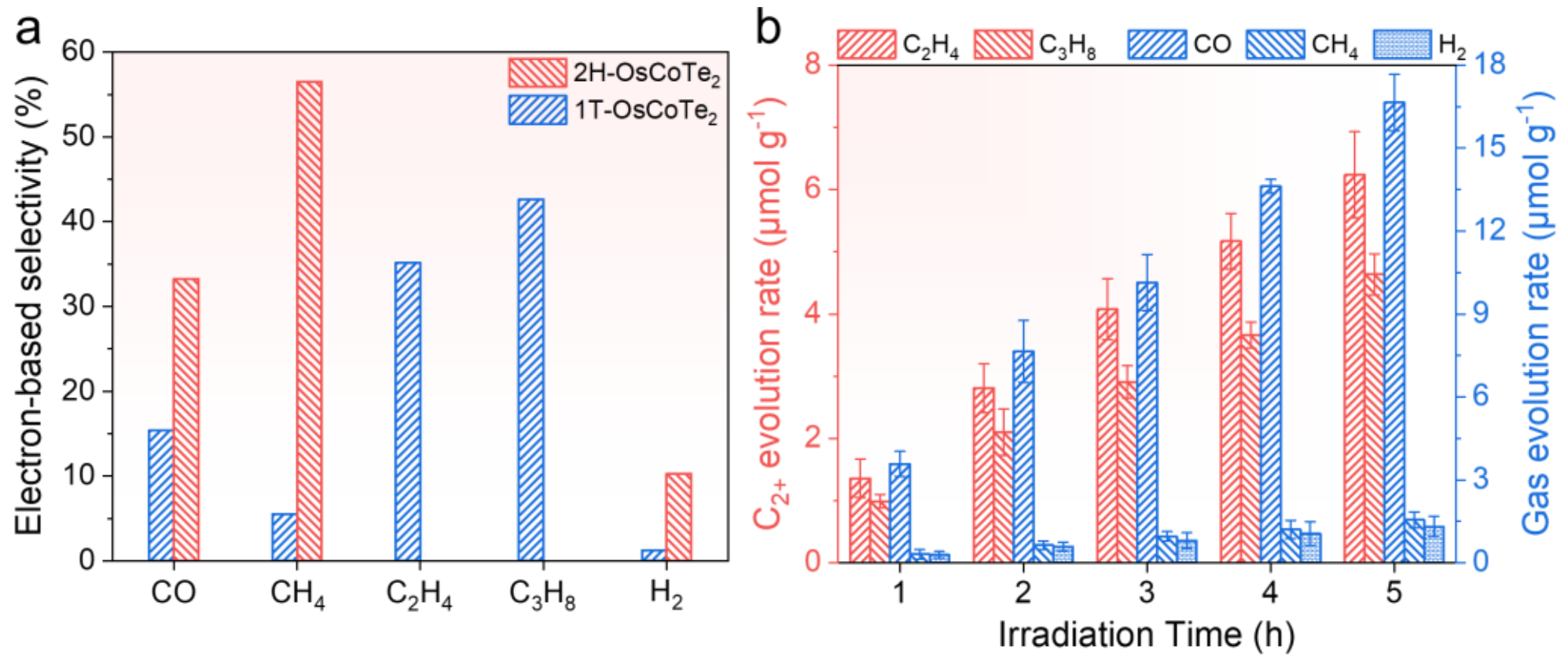

**Fig. S10.** (a) carbon-based product formation rates on 1T-OsCoTe$_2$ and 2H-OsCoTe$_2$ in pure $H_2O$. (b) Photocatalytic carbon-based product evolution as a function of light irradiation time on 1T-OsCoTe$_2$ in pure $H_2O$.

The electron-based selectivity of $C_3H_8$ of 1T-$OsCoTe_2$ in pure $H_2O$ was calculated using:

$$\mathrm{Sel}_{\mathrm{electron}}(C_3H_8) = \left(\frac{20\times n_{(C_3H_8)}}{2\times n_{(H_2)} + 2\times n_{(CO)} + 8\times n_{(CH_4)} + 12\times n_{(C_2H_4)} + 20\times n_{(C_3H_8)}}\right)\times 100\%$$

$$= \left(\frac{20\times 0.99}{2\times 0.29 + 2\times 3.58 + 8\times 0.32 + 12\times 1.36 + 20\times 0.99}\right)\times 100\% \qquad \text{ES1}$$

$$= 42.7\%$$

The electron-based selectivity of $C_{2+}$ ($C_2H_4$ and $C_3H_8$) hydrocarbons of 1T-$OsCoTe_2$ in pure $H_2O$ was calculated using:

$$\mathrm{Sel}_{\mathrm{electron}}(C_{2+}) = \left(\frac{20\times n_{(C_3H_8)} + 12\times n_{(C_2H_4)}}{2\times n_{(H_2)} + 2\times n_{(CO)} + 8\times n_{(CH_4)} + 12\times n_{(C_2H_4)} + 20\times n_{(C_3H_8)}}\right)\times 100\%$$

$$= \left(\frac{20\times 0.99 + 12\times 1.36}{2\times 0.29 + 2\times 3.58 + 8\times 0.32 + 12\times 1.36 + 20\times 0.99}\right)\times 100\% \qquad \text{ES2}$$

$$= 77.8\%$$

The electron-based selectivity of overall $C_{2+}$ ($C_2H_4$ and $C_3H_8$) hydrocarbons over 1T-$OsCoTe_2$ in MeCN aqueous solution is 77.8% and that of $C_3H_8$ is 42.7%.

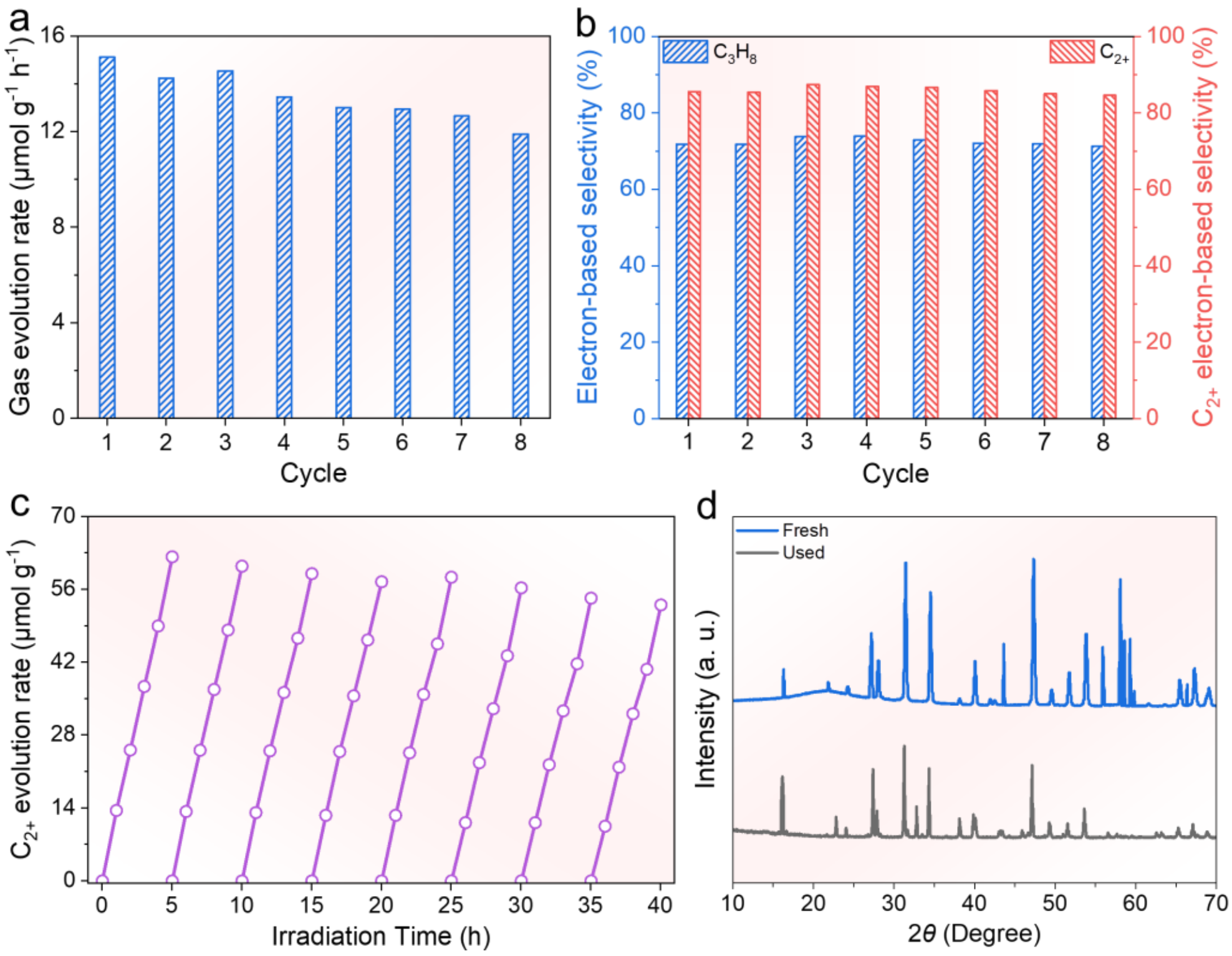

**Fig. S11.** (a) Product formation rate and (b) electron-based selectivity on 1T-$OsCoTe_2$ over eight cycling tests in MeCN aqueous solution. (c) $C_{2+}$ products amounts as a function of light irradiation time for 1T-$OsCoTe_2$ over eight cycling tests in MeCN aqueous solution. (d) XRD patterns of 1T-$OsCoTe_2$ after $CO_2$RR (eight cycles).

The products selectivity of $C_3H_8$ of 1T-$OsCoTe_2$ in MeCN aqueous solution was calculated using:

$$\begin{aligned} \mathrm{Sel_{product}(C_3H_8)} &= \left(\frac{n_{(\mathrm{C_3H_8})}}{n_{(\mathrm{CO})} + n_{(\mathrm{CH_4})} + n_{(\mathrm{C_2H_4})} + n_{(\mathrm{C_3H_8})}}\right) \times 100\% \\ &= \left(\frac{10.28}{14.91+1.78+3.26+10.28}\right) \times 100\% \\ &= 34.2\% \end{aligned} \qquad \text{ES3}$$

The products selectivity of $C_{2+}$ ($C_2H_4$ and $C_3H_8$) of 1T-$OsCoTe_2$ in MeCN aqueous solution was calculated using:

$$\begin{aligned} \mathrm{Sel_{product}(C_3H_8)} &= \left(\frac{n_{(\mathrm{C_3H_8})} + n_{(\mathrm{C_2H_4})}}{n_{(\mathrm{CO})} + n_{(\mathrm{CH_4})} + n_{(\mathrm{C_2H_4})} + n_{(\mathrm{C_3H_8})}}\right) \times 100\% \\ &= \left(\frac{10.28+3.26}{14.91+1.78+3.26+10.28}\right) \times 100\% \\ &= 44.79\% \end{aligned} \qquad \text{ES4}$$

The electron-based selectivity of $C_3H_8$ of 1T-$OsCoTe_2$ in MeCN aqueous solution was calculated using:

$$\begin{aligned} \mathrm{Sel_{electron}(C_3H_8)} &= \left(\frac{20\times n_{(\mathrm{C_3H_8})}}{2\times n_{(\mathrm{CO})} + 8\times n_{(\mathrm{CH_4})} + 12\times n_{(\mathrm{C_2H_4})} + 20\times n_{(\mathrm{C_3H_8})}}\right) \times 100\% \\ &= \left(\frac{20\times 10.28}{2\times 14.91 + 8\times 1.78 + 12\times 3.26 + 20\times 10.28}\right) \times 100\% \\ &= 71.2\% \end{aligned} \qquad \text{ES5}$$

The electron-based selectivity of total $C_{2+}$ hydrocarbons of 1T-$OsCoTe_2$ in MeCN aqueous solution was calculated using:

$$\mathrm{Sel_{electron}}(C_{2+}) = \left(\frac{20\times n_{(C_3H_8)} + 12\times n_{(C_2H_4)}}{2\times n_{(CO)} + 8\times n_{(CH_4)} + 12\times n_{(C_2H_4)} + 20\times n_{(C_3H_8)}}\right)\times 100\%$$

$$= \left(\frac{20\times 10.28 + 12\times 3.26}{2\times 14.91 + 8\times 1.78 + 12\times 3.26 + 20\times 10.28}\right)\times 100\% \qquad \text{ES6}$$

$$= 84.75\%$$

The electron-based selectivity of overall $C_{2+}$ ($C_2H_4$ and $C_3H_8$) hydrocarbons of 2H-$OsCoTe_2$ in pure $H_2O$ decreases to 84.75% and that of $C_3H_8$ decreases to 71.2%.

**Table S2.** The calculated *AQE* and $IQE_{cr}$ of 1T-$OsCoTe_2$ at different wavelengths.

| Wavelength (nm) | *N* | Products production (μmol) | | | | *AQE* (%) | *A* | $IQE_{cr}$ (%) |
|---|---|---|---|---|---|---|---|---|
| | | $C_3H_8$ | $C_2H_4$ | $CH_4$ | CO | | | |
| 380 | $8.29\times10^{19}$ | 1.731 | 0.932 | 0.791 | 7.983 | 49.44 | 0.905 | 54.63 |
| 400 | $9.89\times10^{19}$ | 1.381 | 0.881 | 0.632 | 5.65 | 21.84 | 0.852 | 25.63 |
| 420 | $1.09\times10^{20}$ | 0.589 | 0.16 | 0.09 | 0.75 | 8.81 | 0.813 | 10.84 |
| 440 | $1.03\times10^{20}$ | 0.467 | 0.134 | 0.072 | 0.62 | 7.46 | 0.786 | 9.49 |
| 460 | $1.07\times10^{20}$ | 0.333 | 0.117 | 0.083 | 0.42 | 5.43 | 0.763 | 7.12 |
| 480 | $1.02\times10^{20}$ | 0.233 | 0.071 | 0.051 | 0.34 | 3.91 | 0.745 | 5.25 |
| 500 | $1.29\times10^{20}$ | 0.162 | 0.052 | 0.026 | 0.18 | 2.06 | 0.736 | 2.80 |
| 520 | $1.52\times10^{20}$ | 0.067 | 0.023 | 0.071 | 0.08 | 0.93 | 0.734 | 1.27 |

**Apparent quantum efficiency (*AQE*).** The *AQE* was determined under the above photocatalytic reaction conditions except that Xe lamp was used as the irradiation source. The wavelength-dependent *AQE* was measured under the same photocatalytic reaction condition, except for the monochromatic light wavelengths (380, 400, 420, 440, 460, 480, 500, and 520 nm). All the experiments were repeated at least 3 times in parallel to obtain an average value. At the same time, the *AQE* and corresponding *IQE* ($IQE_{cr}$) were roughly calculated from the equation: [2, 3]

$$AQE = \frac{\text{the number of reacted electrons}}{\text{the number of incident photons}} \times 100\%$$

$$= \frac{\text{the number of reacted electrons}}{N} \times 100\% \qquad \text{ES7}$$

$$= \frac{\text{the number of reacted electrons}}{\frac{E\lambda}{hc}} \times 100\%$$

$$IQE_{\text{cr}} = \frac{AQE}{A} \times 100\% \qquad \text{ES8}$$

where $A$ and $N$ is the absorption of sample solution and number of incident photons.

Taking 1T-$OsCoTe_2$ at $\lambda$ = 420 nm as an example:

$$N = \frac{E\lambda}{hc} = \frac{1.43 \times 10^{-2} \times 3600 \times 420 \times 10^{?9}}{6.626 \times 10^{-34} \times 3 \times 10^{8}} = 1.09 \times 10^{20}$$

$$AQE_{(\text{C}_3\text{H}_8)} = \frac{20 \times 0.589 \times 10^{-6} \times 6.02 \times 10^{23}}{1.088 \times 10^{20}} \times 100\% = 6.52\%$$

$$AQE_{(\text{C}_2\text{H}_4)} = \frac{12 \times 0.16 \times 10^{-6} \times 6.02 \times 10^{23}}{1.088 \times 10^{20}} \times 100\% = 1.06\%$$

$$AQE_{(\text{CH}_4)} = \frac{8 \times 0.09 \times 10^{-6} \times 6.02 \times 10^{23}}{1.088 \times 10^{20}} \times 100\% = 0.4\%$$

$$AQE_{(\text{CO})} = \frac{2 \times 0.75 \times 10^{-6} \times 6.02 \times 10^{23}}{1.088 \times 10^{20}} \times 100\% = 0.83\%$$

$$\text{Total } AQE = AQE_{(\text{C}_3\text{H}_8)} + AQE_{(\text{C}_2\text{H}_4)} + AQE_{(\text{CH}_4)} + AQE_{(\text{CO})} = 8.81\%$$

$$IQE_{\text{cr}} = \frac{\text{Total } AQE}{A} \times 100\% = \frac{8.81\%}{0.813} \times 100\% = 10.84\%$$

**Table S3.** A summary of photocatalytic $CO_2$ reduction performances by the reported Co-based catalysts using $H_2O$ and MeCN as the electron donor.

| Catalysts | Reaction medium | Rate (μmol $h^{-1}g^{-1}$) | Ref. |
|---|---|---|---|
| FeTCP-OH-Co | $H_2O$ | 17.72 (HCOOH) | [4] |
| Co-ZIF-9/$TiO_2$ | $H_2O$ | 17.6 (CO)<br>2.0 ($CH_4$) | [5] |
| BV/CoDAC | $H_2O$ | 19.7 ($CH_4$) | [6] |
| $FeCoS_2$ | $H_2O$ | 20.2 (CO) | [7] |
| $CsPbBr_3$/$Cs_4PbBr_6$@Co | $H_2O$ | 11.95 (CO) | [8] |
| Co-$NiS_2$ | $H_2O$ | 2.35 ($C_2H_4$) | [9] |
| CoPcPDA-CMP NSs | 1 mL $H_2O$ | 14.3 (CO) | [10] |
| $CsPbBr_3$/ZIF | $H_2O$ | 4.09 (CO)<br>2.35 ($CH_4$) | [11] |
| NiCo-$TiO_2$ | $H_2O$+0.1 M $Na_2SO_3$+0.2 M CsOH | 22.6 (Acetate) | [12] |
| Ni-Co/g-$C_3N_4$ | 5 mL $H_2O$ | 0.45 ($CH_4$) | [13] |
| 1T-$OsCoTe_2$<br>2H-$OsCoTe_2$ | MeCN+$H_2O$ | 10.24 ($C_3H_8$)<br>7.72 ($C_3H_8$) | This work |

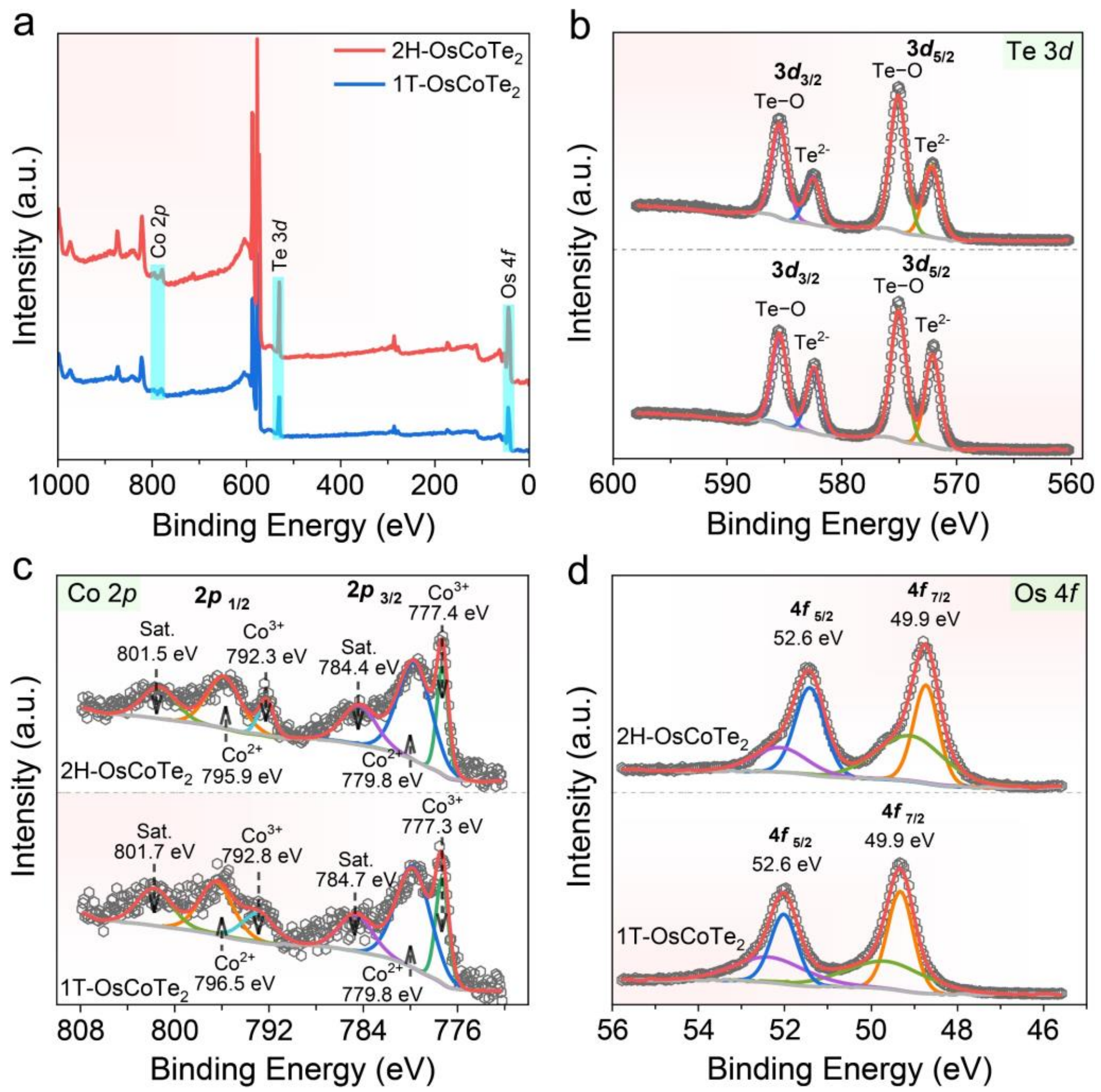

**Fig. S12.** (a) XPS full-scan survey, (b) XPS Te 3*d*, (c) Co 2*p*, (d) Os 4*f* spectra of 1T-$OsCoTe_2$ and 2H-$OsCoTe_2$ after $CO_2$RR (eight cycles).

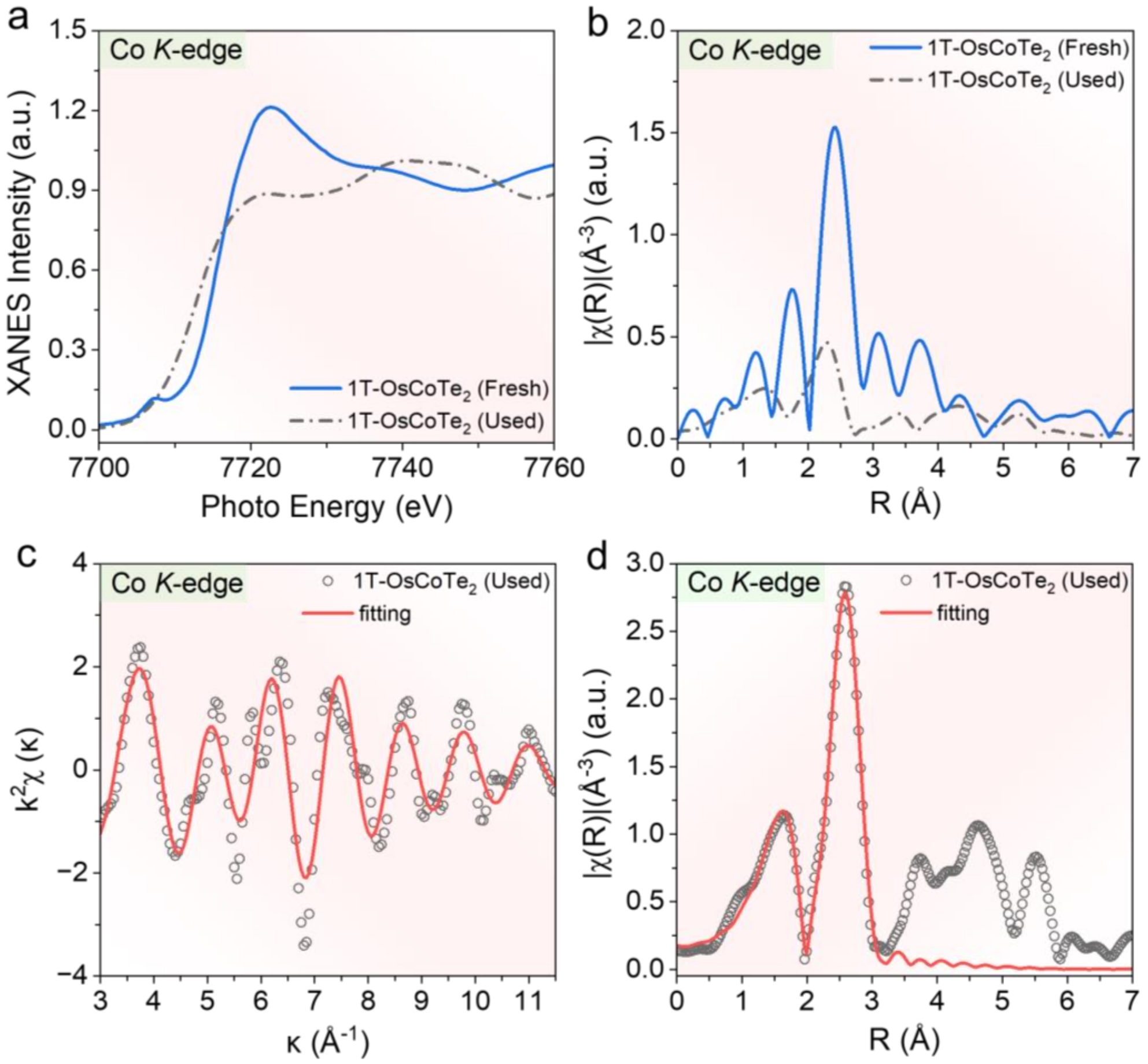

**Fig. S13.** (a) Co *K*-edge XANES spectra of different samples. (b) Co *K*-edge EXAFS spectra in R-pace. EXFAS fitting curves for in (c) k-pace and (d) R-pace of 1T-$OsCoTe_2$ (Used).

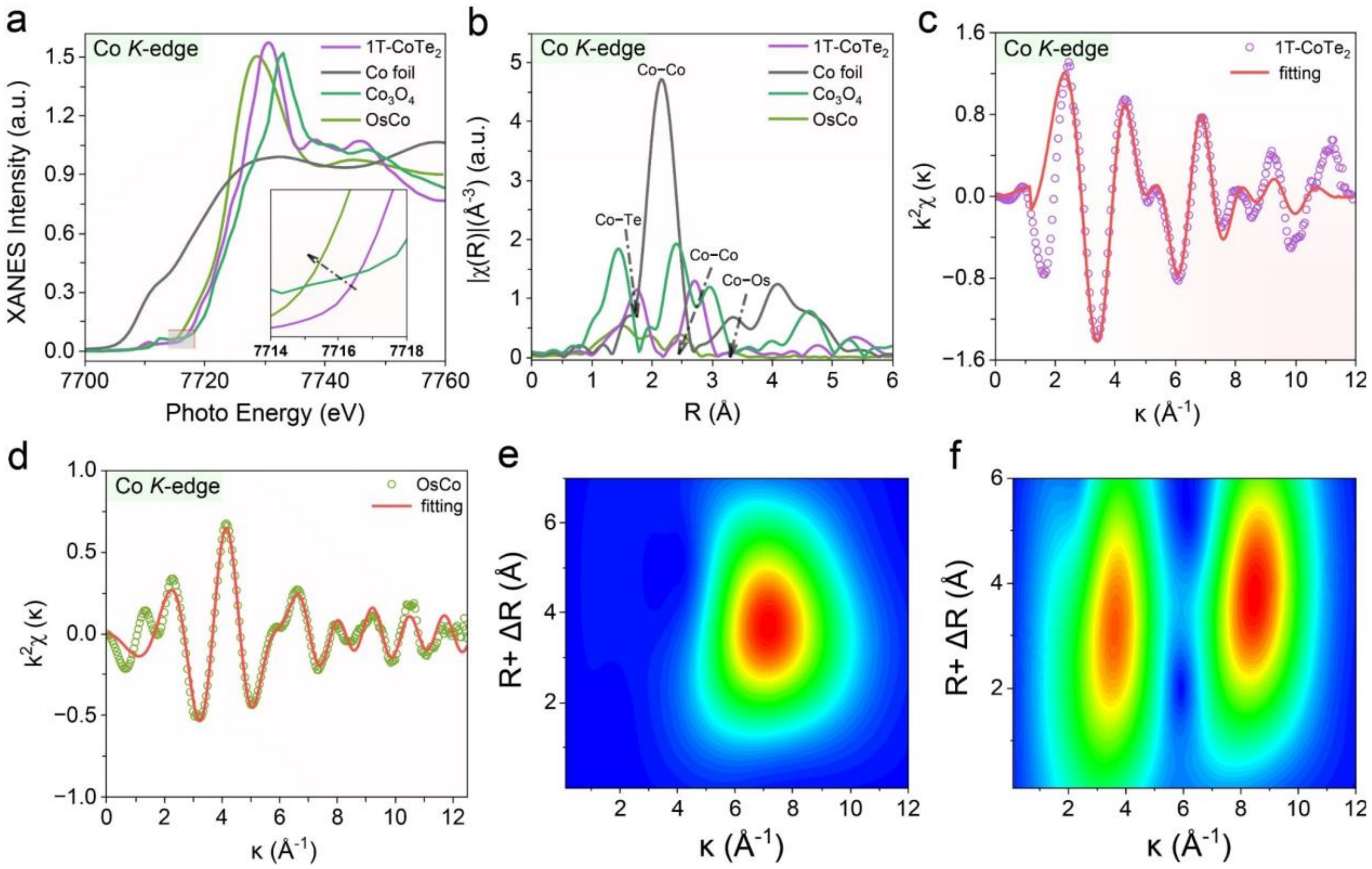

**Fig. S14.** (a) Co *K*-edge XANES spectra of different samples. (b) Co *K*-edge extended EXAFS spectra in R-pace. EXFAS fitting curves for (c) 1T-$CoTe_2$ and (d) OsCo. WT-EXAFS spectra in discriminating radial distance and k-pace resolution of (e) 1T-$CoTe_2$ and (f) OsCo.

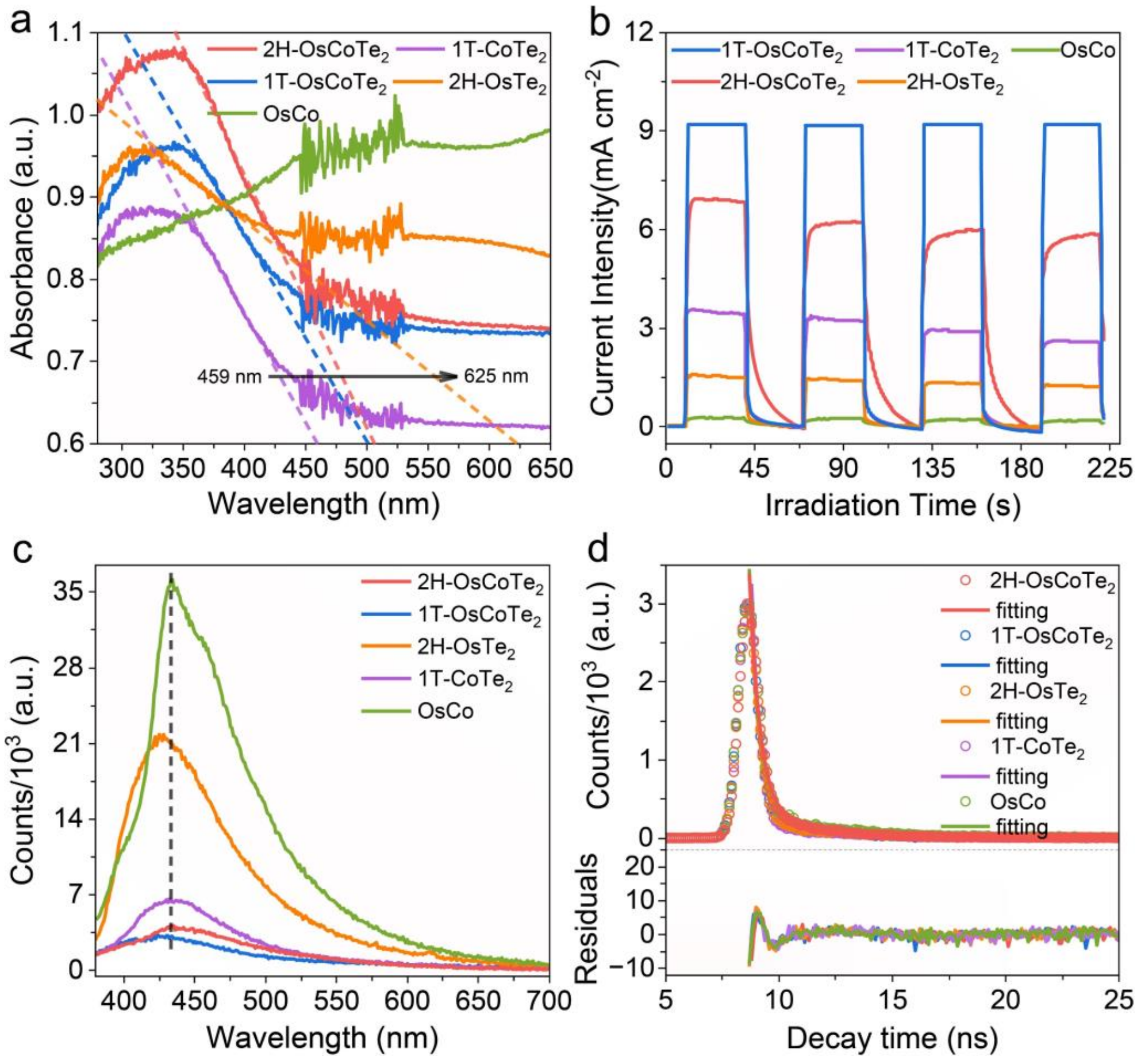

**Fig. S15.** (a) UV-vis DRS, (b) transient photocurrents (1.0 M KOH, λ ≥ 420 nm), (c) PL, and (d) TRPL spectra of $1T-OsCoTe_2$, $2H-OsCoTe_2$, $2H-OsTe_2$, $1T-CoTe_2$, and OsCo, respectively.

**Table S5** The average fluorescence lifetime of samples.

| Samples | $\tau_1$ (ns) | $B_1$ (%) | $\tau_2$ (ns) | $B_2$ (%) | $\tau_3$ (ns) | $B_3$ (%) | Ave. $\tau$(ns) |
|---|---|---|---|---|---|---|---|
| $1T-OsCoTe_2$ | 0.4215 | 71.61 | 3.1925 | 25.46 | 13.3792 | 2.93 | 1.50 |
| $2H-OsCoTe_2$ | 0.4838 | 66.25 | 2.7166 | 31.24 | 15.0095 | 2.51 | 1.5461 |
| $1T-CoTe_2$ | 0.3623 | 69.53 | 3.6829 | 26.37 | 16.0284 | 4.10 | 1.88 |
| $2H-OsTe_2$ | 0.4204 | 68.44 | 3.6939 | 25.89 | 14.6650 | 5.67 | 2.0753 |
| OsCo | 0.4582 | 58.85 | 3.3855 | 31.62 | 11.1796 | 9.53 | 2.4053 |

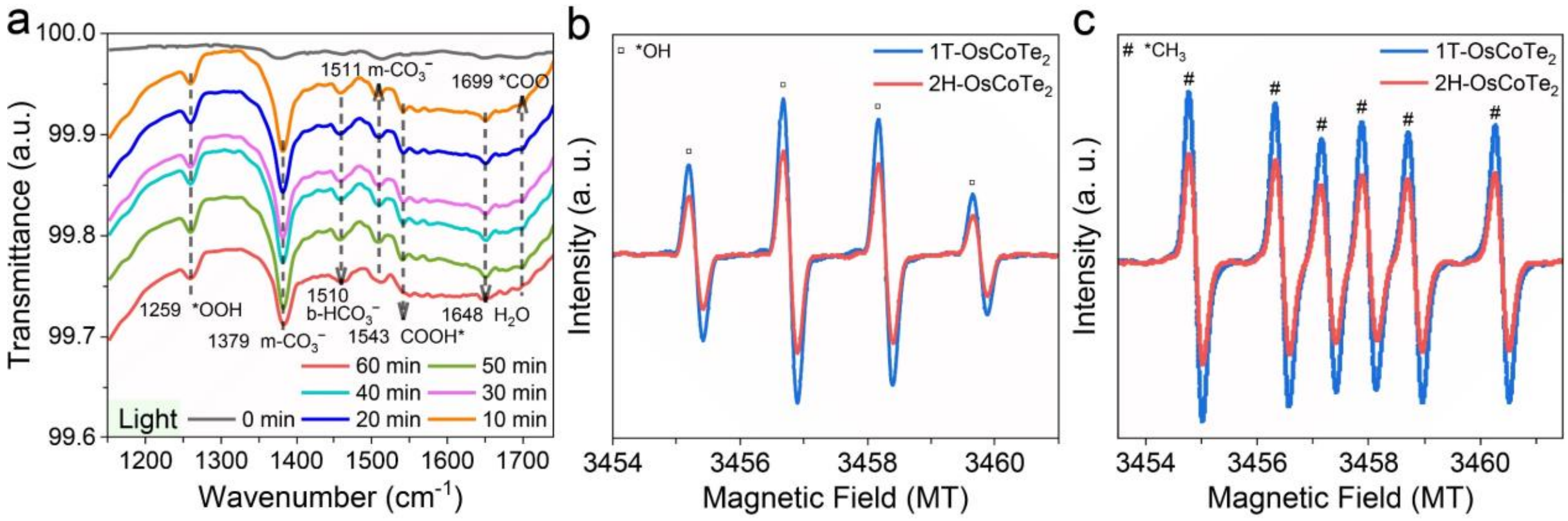

**Fig. S16.** (a) In-situ DRIFT spectra for light-driven $CO_2$ conversion over 2H-$OsCoTe_2$. In-situ ESR spectra of (b) $H_2O$ (5 mL $H_2O$+3 μL DMPO) and (c) $CO_2$+isopropyl alcohol mixture ($CO_2$+5 mL isopropyl alcohol+3 μL DMPO) under light illumination in the presence of DMPO over 1T-$OsCoTe_2$ and 2H-$OsCoTe_2$ at 298 K.

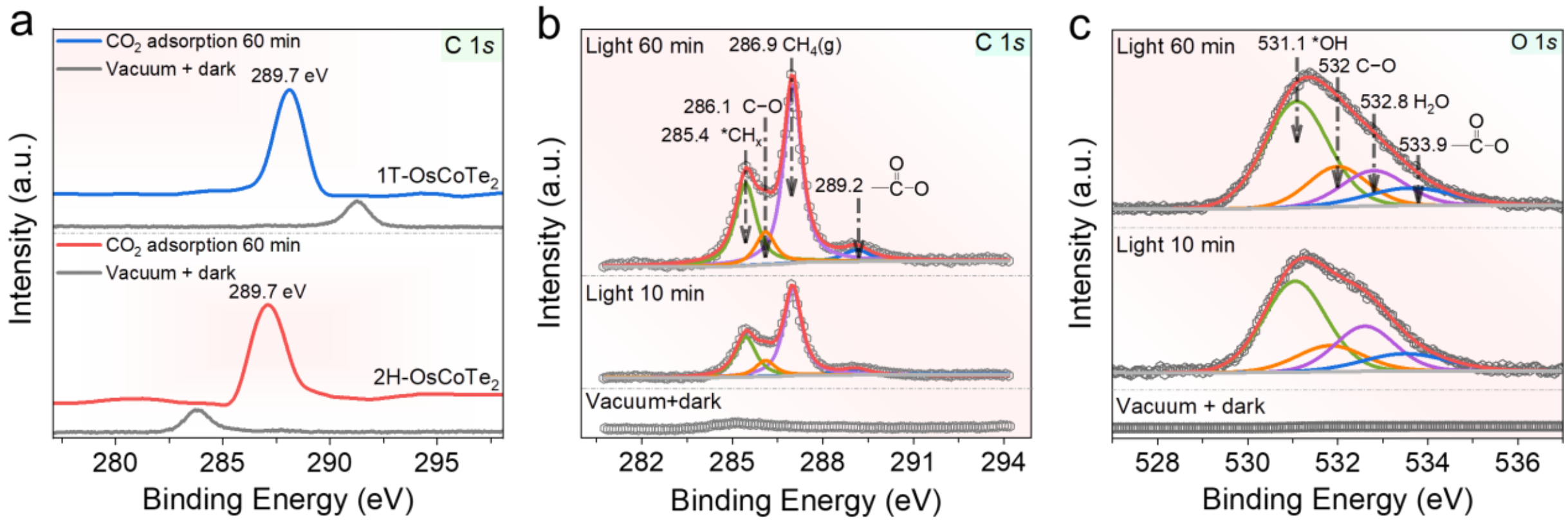

**Fig. S17.** High-resolution C 1*s* XPS spectra of (a) 1T-$OsCoTe_2$ and 2H-$OsCoTe_2$ under vacuum condition and after $CO_2$ adsorption for 60 min. In situ NAP-XPS results of high-resolution (b) C 1*s* and (c) O 1*s* spectra over 2H-$OsCoTe_2$.

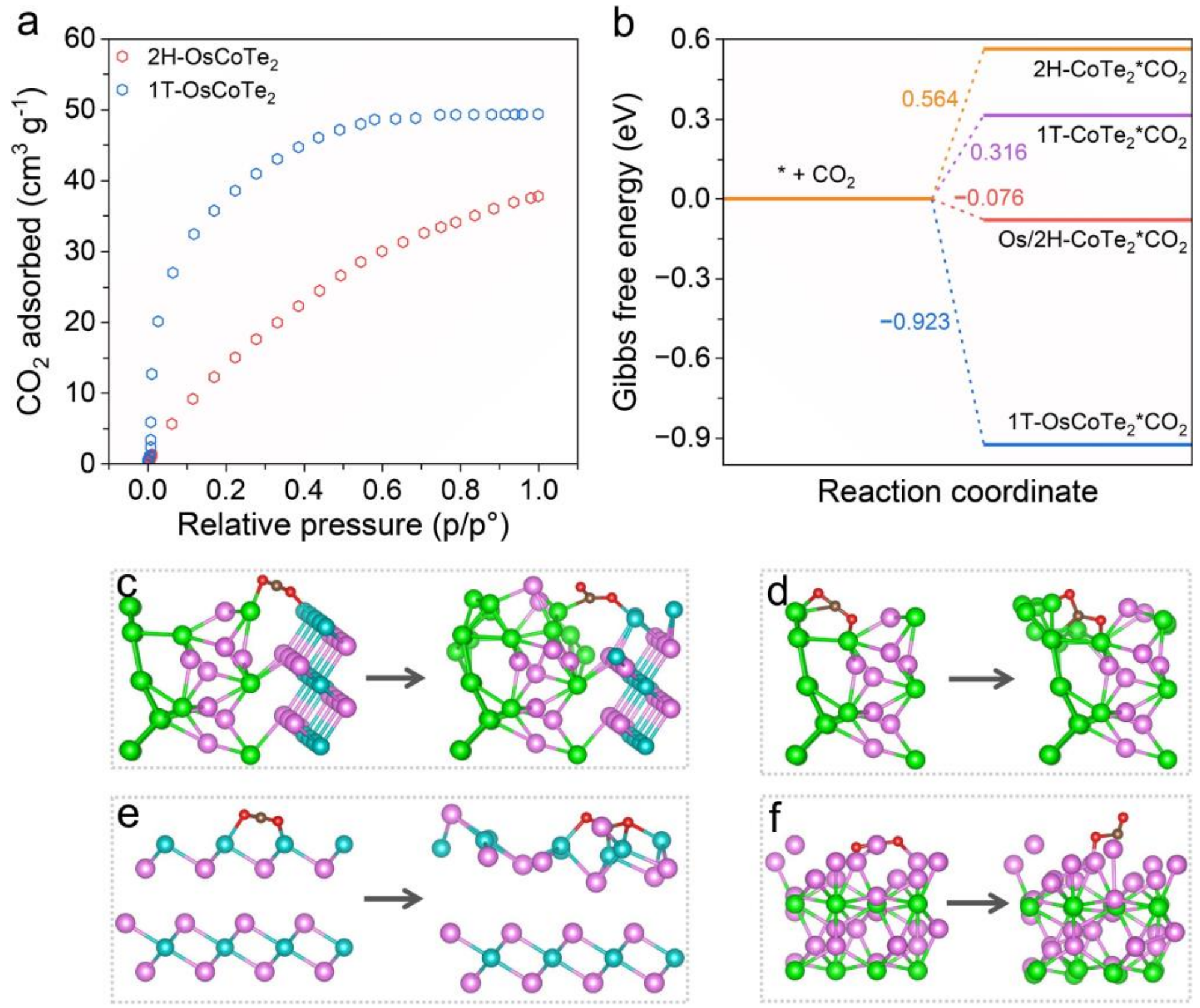

**Fig. S18.** (a) $CO_2$ adsorption isotherms of 1T-$OsCoTe_2$ and 2H-$OsCoTe_2$ at 298 K. (b) The Gibbs free energy changes of $CO_2$ activation, and the corresponding adsorption configurations of *$CO_2$ on (c) 1T-$OsCoTe_2$, (d) Os/2H-$OsTe_2$, (e) 1T-$CoTe_2$, and (f) 2H-$OsTe_2$ (side view and color codes in the atomic models: green (Os), blue (Co), pink (Te), red (O) and brown (C).

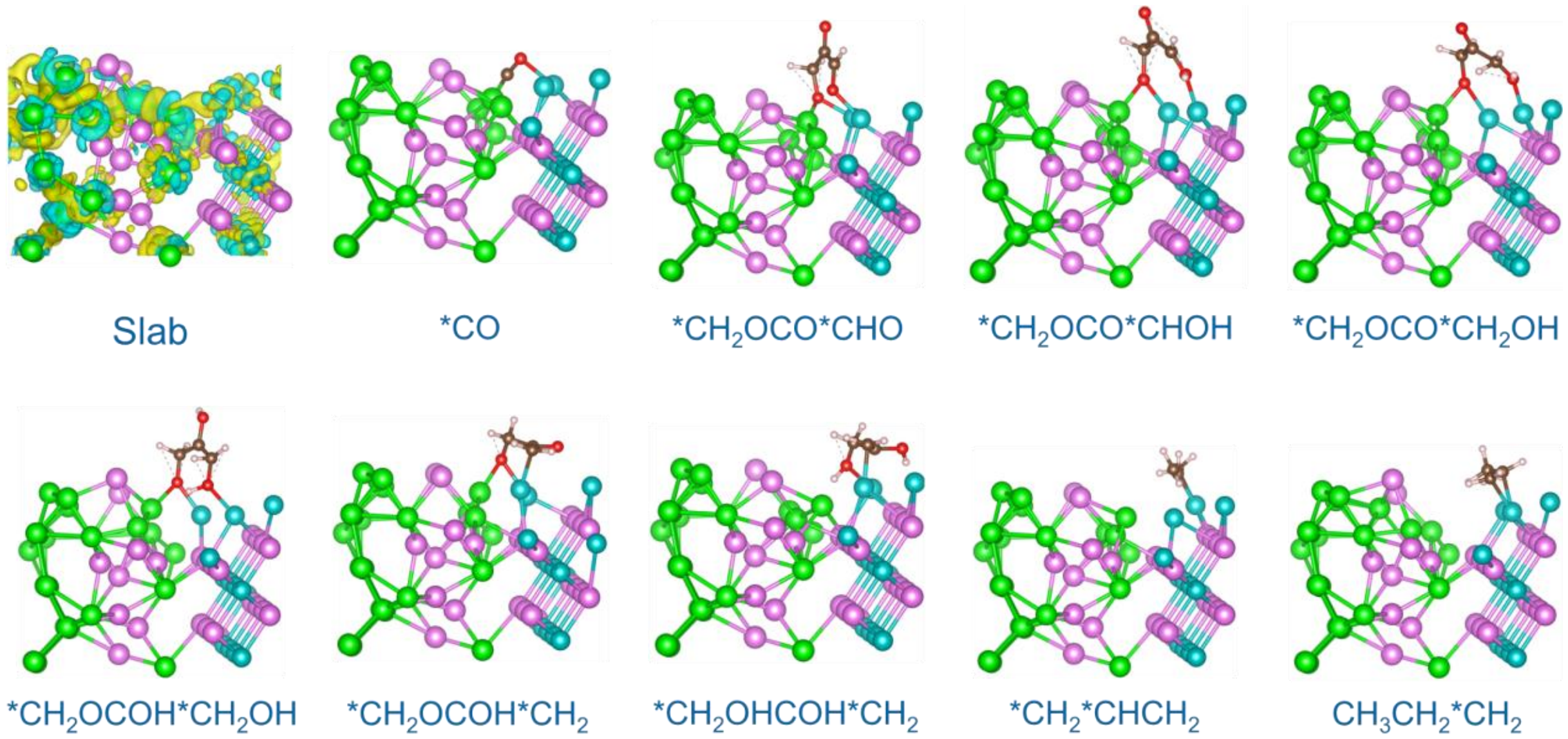

**Fig. S19.** The model diagrams of the products on 1T-$OsCoTe_2$.

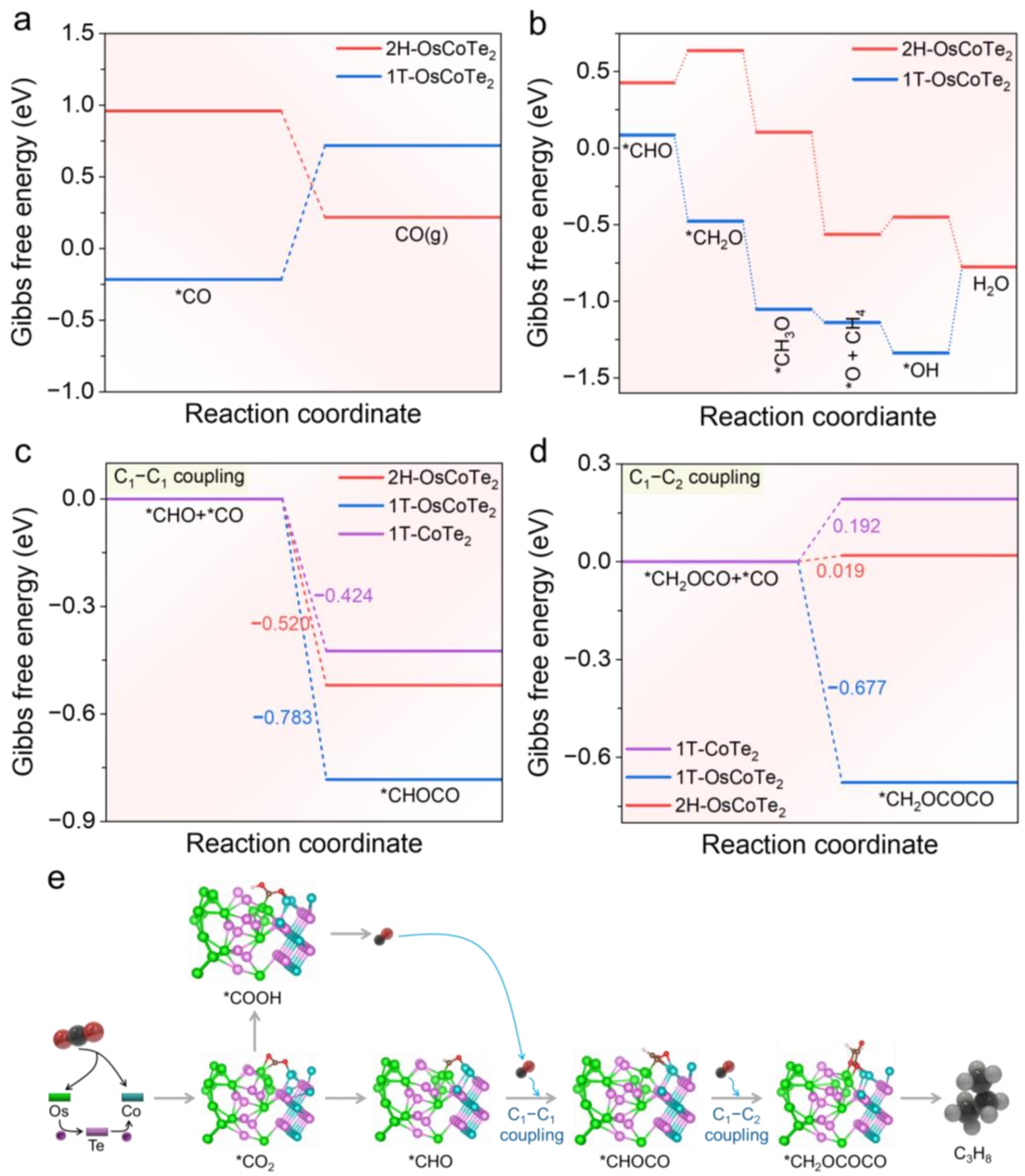

**Fig. S20.** Gibbs free energy for steps in photocatalytic (a) *CO-to-CO and (b) $CO_2$-to-$CH_4$ conversion on 1T-$OsCoTe_2$ and 2H-$OsCoTe_2$. Comparison for the Gibbs free energy changes of (c) $C_1$—$C_1$ coupling and (d) $C_1$—$C_2$ coupling on 1T-$OsCoTe_2$. (e) Illustration of the roles of the Co and Os of $CO_2$ to $C_3H_8$ on 1T-$OsCoTe_2$.

## References

[1] B. Ravel, M. Newville, ATHENA, ARTEMIS, HEPHAESTUS: data analysis for X-ray absorption spectroscopy using IFEFFIT, J. Synchrotron Radiat., 12 (2005) 537-541.

[2] C. Ding, X. Lu, B. Tao, L. Yang, X. Xu, L. Tang, H. Chi, Y. Yang, D.M. Meira, L. Wang, X. Zhu, S. Li, Y. Zhou, Z. Zou, Interlayer Spacing Regulation by Single-Atom Indium$^{\delta+}$–$N_4$ on

Carbon Nitride for Boosting $CO_2$/CO Photo-Conversion, Adv. Funct. Mater., 33 (2023) 2302824.

[3] Z. Xie, S. Xu, L. Li, S. Gong, X. Wu, D. Xu, B. Mao, T. Zhou, M. Chen, X. Wang, W. Shi, S. Song, Well-defined diatomic catalysis for photosynthesis of $C_2H_4$ from $CO_2$, Nat. Commun., 15 (2024) 2422.

[4] E.-X. Chen, M. Qiu, Y.-F. Zhang, L. He, Y.-Y. Sun, H.-L. Zheng, X. Wu, J. Zhang, Q. Lin, Energy Band Alignment and Redox-Active Sites in Metalloporphyrin-Spaced Metal-Catechol Frameworks for Enhanced $CO_2$ Photoreduction, Angew. Chem. Int. Ed., 61 (2022) e202111622.

[5] S. Yan, S. Ouyang, H. Xu, M. Zhao, X. Zhang, J. Ye, Co-ZIF-9/$TiO_2$ nanostructure for superior $CO_2$ photoreduction activity, J. Mater. Chem. A, 4 (2016) 15126-15133.

[6] K. Zhu, Q. Zhu, M. Jiang, Y. Zhang, Z. Shao, Z. Geng, X. Wang, H. Zeng, X. Wu, W. Zhang, K. Huang, S. Feng, Modulating Ti $t_{2g}$ Orbital Occupancy in a Cu/$TiO_2$ Composite for Selective Photocatalytic $CO_2$ Reduction to CO, Angew. Chem. Int. Ed., 61 (2022) e202207600.

[7] Y. Wu, Q. Chen, J. Zhu, K. Zheng, M. Wu, M. Fan, W. Yan, J. Hu, J. Zhu, Y. Pan, X. Jiao, Y. Sun, Y. Xie, Selective $CO_2$-to-$C_2H_4$ Photoconversion Enabled by Oxygen-Mediated Triatomic Sites in Partially Oxidized Bimetallic Sulfide, Angew. Chem. Int. Ed., 62 (2023) e202301075.

[8] Y.-F. Mu, W. Zhang, X.-X. Guo, G.-X. Dong, M. Zhang, T.-B. Lu, Water-Tolerant Lead Halide Perovskite Nanocrystals as Efficient Photocatalysts for Visible-Light-Driven $CO_2$ Reduction in Pure Water, ChemSusChem, 12 (2019) 4769-4774.

[9] W. Shao, X. Li, J. Zhu, X. Zu, L. Liang, J. Hu, Y. Pan, J. Zhu, W. Yan, Y. Sun, Y. Xie, Metal$^{n+}$-Metal$^{\delta+}$ pair sites steer C—C coupling for selective $CO_2$ photoreduction to $C_2$ hydrocarbons, Nano Res., 15 (2022) 1882-1891.

[10] Q. Zhi, J. Zhou, W. Liu, L. Gong, W. Liu, H. Liu, K. Wang, J. Jiang, Covalent Microporous Polymer Nanosheets for Efficient Photocatalytic $CO_2$ Conversion with $H_2O$, Small, 18 (2022) 2201314.

[11] L.-Y. Wu, Y.-F. Mu, X.-X. Guo, W. Zhang, Z.-M. Zhang, M. Zhang, T.-B. Lu, Encapsulating Perovskite Quantum Dots in Iron-Based Metal–Organic Frameworks (MOFs) for Efficient Photocatalytic $CO_2$ Reduction, Angew. Chem. Int. Ed., 58 (2019) 9491-9495.

[12] G. Jia, M. Sun, Y. Wang, Y. Shi, L. Zhang, X. Cui, B. Huang, J. C. Yu, Asymmetric Coupled Dual-Atom Sites for Selective Photoreduction of Carbon Dioxide to Acetic Acid, Adv. Funct. Mater., 32 (2022) 2206817.

[13] X. Zhang, X. Zhang, W. Ali, X. Chen, K. Hu, Z. Li, Y. Qu, L. Bai, Y. Gao, L. Jing, Improved Photoactivities of Large-surface-area g-$C_3N_4$ for $CO_2$ Conversion by Controllably

Introducing Co- and Ni-Species to Effectively Modulate Photogenerated Charges, ChemCatChem, 11 (2019) 6282-6287.